\documentclass[aps,pra,twocolumn,titlepage,superscriptaddress,floatfix]{revtex4-1}
\usepackage[english]{babel}
\usepackage[utf8x]{inputenc}
\usepackage[T1]{fontenc}
\usepackage{lipsum}
\usepackage{bm}
\usepackage{dsfont}
\usepackage{amsmath}
\usepackage{amsmath,esint}
\usepackage{float}

\usepackage{physics}
\usepackage{amsmath}
\usepackage{graphicx}
\usepackage[colorinlistoftodos]{todonotes}
\usepackage[colorlinks=true, allcolors=blue]{hyperref}

\newcommand{\jcmt}[2]{\hspace{0in}#2}

\makeatletter
\def\Arg{\mathop{\operator@font Arg}\nolimits}
\makeatother

\begin{document}
\title{Probing the Berry Curvature and Fermi Arcs of a Weyl Circuit}

\author{Yuehui Lu}
\affiliation{The James Franck Institute and Department of Physics, University of Chicago, Chicago, Illinois 60637, USA}
\author{Ningyuan Jia}
\affiliation{The James Franck Institute and Department of Physics, University of Chicago, Chicago, Illinois 60637, USA}
\author{Lin Su}
\affiliation{The James Franck Institute and Department of Physics, University of Chicago, Chicago, Illinois 60637, USA}
\author{Clai Owens}
\affiliation{The James Franck Institute and Department of Physics, University of Chicago, Chicago, Illinois 60637, USA}
\author{Gediminas Juzeliūnas}
\affiliation{Institute of Theoretical Physics and Astronomy, Vilnius University, Vilnius, Lithuania}
\author{David I. Schuster}
\affiliation{The James Franck Institute and Department of Physics, University of Chicago, Chicago, Illinois 60637, USA}
\author{Jonathan Simon}
\email{simonjon@uchicago.edu}
\affiliation{The James Franck Institute and Department of Physics, University of Chicago, Chicago, Illinois 60637, USA}

\date{\today}
\begin{abstract}
The Weyl particle is the massless fermionic cousin of the photon~\cite{armitage2018weyl}. While no fundamental Weyl particles have been identified, they arise in condensed matter ~\cite{xu2015discovery,lv2015experimental,hasan2017discovery} and meta-material~\cite{lu2015experimental,noh2017experimental} systems, where their spinor nature imposes topological constraints on low-energy dispersion and surface properties. Here we demonstrate a topological circuit with Weyl dispersion at low-momentum, realizing a 3D lattice that behaves as a half-flux Hofstadter model in all principal planes~\cite{dubvcek2015weyl}. The circuit platform~\cite{ningyuan2015time} provides access to the complete complex-valued spin-texture of all bulk- and surface- states, thereby revealing not only the presence of Weyl points and the Fermi arcs that connect their surface-projections, but also, for the first time, the Berry curvature distribution through the Brillouin zone and the associated quantized Chiral charge of the Weyl points. This work opens a path to exploration of interacting Weyl physics~\cite{wei2012excitonic} in superconducting circuits~\cite{wallraff2004strong}, as well as studies of how manifold topology impacts band topology in three dimensions~\cite{weststrom2017designer}.
\end{abstract}

\maketitle

Creating and probing particles with topologically non-trivial dispersion is a growing endeavor with benefits from exploration of exotic emergent phenomenology in manybody physics~\cite{bloch2008many,carusotto2013quantum,goldman2014light}, to next-generation technologies (e.g., waveguides~\cite{hafezi2011robust} and circulators~\cite{mahoney2017chip}). In two dimensions, successes range from synthetic realizations of graphene~\cite{tarruell2012creating,gomes2012designer} and Haldane's model~\cite{jotzu2014experimental}, to spin-orbit coupling~\cite{lin2011spin,Wang2018} and gauge fields~\cite{lin2009synthetic, hafezi2013imaging, wang2009observation, rechtsman2013photonic, ningyuan2015time, owens2018quarter, schine2016synthetic, susstrunk2015observation, nash2015topological, tai2017microscopy}. Of particular interest are implementations compatible with strong interactions between individual quantized excitations, where extensions to the strongly correlated regime are possible. Candidate platforms include ultracold atoms in optical lattices~\cite{lewenstein2007ultracold}, microwave photons in superconducting circuits~\cite{wallraff2004strong}, and Rydberg-dressed photons~\cite{peyronel2012quantum} in optical resonators~\cite{jia2017strongly}.

Recently, there has been growing interest in exploring the properties of three-dimensional quasi-particles, with a particular focus on Weyl particles, as they have \emph{not} been observed in nature. With a Hamiltonian of the form $H\sim \bm{\sigma}\cdot\bm{p}$, these massless particles have a linear dispersion $E\propto \pm|\bm{p}|$, and are \emph{chiral}, meaning that their eigenstates exhibit momentum-dependent spin-texture, with a spin-momentum aligned high-energy branch, and an anti-aligned low-energy branch: the momentum acts as an effective \emph{Zeeman} field for the spin in a 3D analog of the Dirac fermion~\cite{tarruell2012creating,gomes2012designer}. The additional symmetries of a lattice system prevent Weyl dispersion over the full Brillouin zone (BZ), restricting the behavior to the vicinity of ``Weyl points''.

Weyl dispersion has recently been observed in both optical and microwave meta-materials: A type-I Weyl node, corresponding to a point-like Fermi-surface with linear dispersion, was imaged in the projected dispersion relation of a gyroid microwave material via angle-resolved transmission~\cite{lu2015experimental}, and the robustness of the surface states demonstrated through introduction of local defects~\cite{chen2016photonic}; a Type-II Weyl node, a highly tilted Weyl dispersion (see Fig. 5 of ref.~\cite{armitage2018}) where particle- and hole- pockets touch at a point~\cite{soluyanov2015type}, was observed through conical diffraction, along with Fermi-arc-like surface states in an array of laser-written waveguides~\cite{noh2017experimental} and hyperbolic microwave meta-materials~\cite{yang2017direct}. Such a spinful 3D model \emph{must} exhibit non-trivial spin-texture in the vicinity of a linear dispersion point (Weyl point), but to our knowledge, no system (prior to this work) has managed to directly measure the Berry curvature associated with this spin texture.

By extending our $Z_2$ topological circuit~\cite{ningyuan2015time,albert2015topological} into the third dimension~\cite{lee2017topolectrical, luo2018topological,yang2018realization}, we provide the first experimental realization of Weyl particles in a circuit. In a 3D array of low-loss lumped circuit elements, we implement a cubic lattice with $\pi$-flux per plaquette in all principal planes~\cite{dubvcek2015weyl}, realizing a Weyl band structure. Thanks to the exquisite control afforded by the circuit platform, we reveal the system's complex response with site-, energy- and spin- resolved microscopy. With the flexibility of non-local couplings between system edges, we impose periodic boundary conditions on some or all surfaces to probe both bulk and surface physics. We are thus able to measure the full spin-resolved band-structure of the meta-material. We extract the Berry curvature from the band-structure and ascertain that the Weyl points are indeed quantized sources and sinks of Berry flux called ``chiral charges.'' Finally, we perform a full reconstruction of the surface states vs. momentum \& energy and observe that the surface-projections of the the Weyl points are indeed connected by Fermi arcs.

\begin{figure*}
\centering
\includegraphics[width=\textwidth]{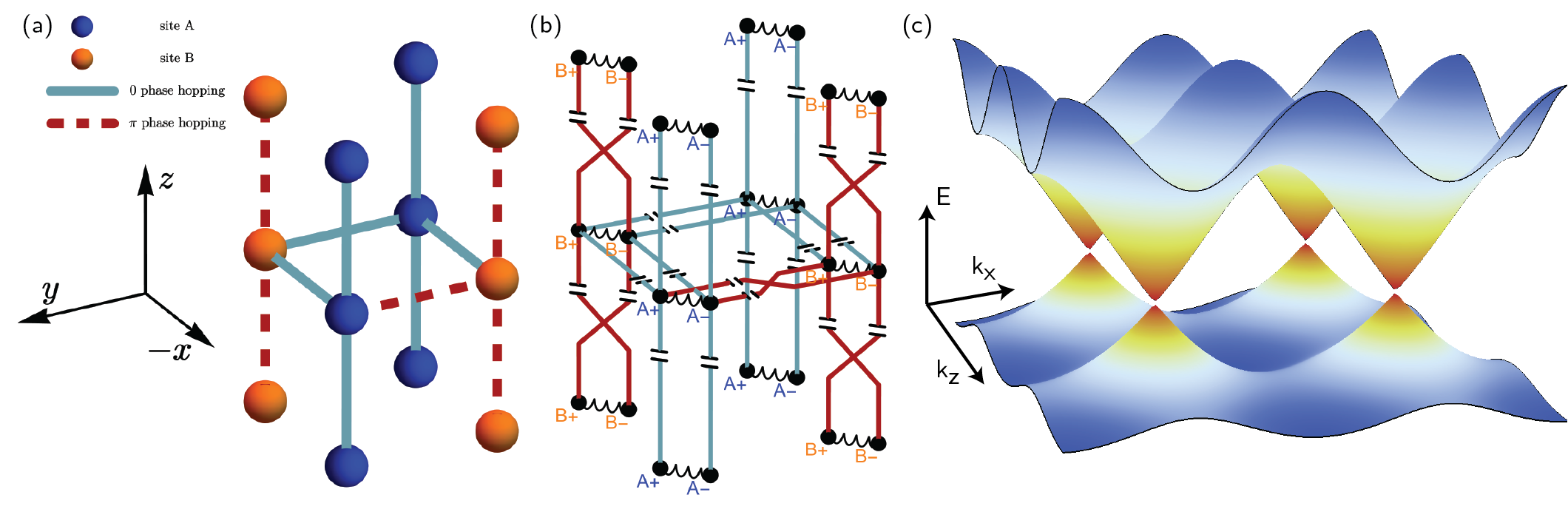}
\caption{\label{fig:setup}\textbf{A Weyl Circuit.} \textbf{(a)} Topology of the tunneling connectivity of a minimal lattice model exhibiting Weyl points. The unit cell consists of two sites, with \textbf{A}-sites shown in blue, and \textbf{B}-sites shown in orange. Zero-phase tunneling is represented as cyan solid line, and $\pi$-phase tunneling as red dashed line. The lattice vectors are $\hat{z}$, $\hat{u}\equiv\frac{1}{\sqrt{2}}(\hat{x}+\hat{y})$, and $\hat{v}\equiv\frac{1}{\sqrt{2}}(\hat{x}-\hat{y})$. \textbf{(b)} To realize such a lattice with circuit components, all lattice-sites are replaced by inductors, and all tunnel-couplings by a pair of capacitors. A zero-phase tunnel-coupling capacitively connects the positive end of one inductor to the positive end of its neighbor, and the negative end to the negative end of the neighbor. A $\pi$-phase tunnel-connection capacitively couples the positive end of an inductor to the \emph{negative} end of its neighbor, and vice-versa. \textbf{(c)} Shows numerically calculated band structure at $k_y=0$; apparent are four Weyl points in the first Brillouin zone.
}
\end{figure*}

\section{Engineering a Weyl Circuit}
The Weyl model that we realize in our circuit is analogous to a recent cold-atom proposal~\cite{dubvcek2015weyl} that may be viewed as either  a cubic lattice with $\pi$-flux penetrating each plaquette in each principal plane ($x-y$, $x-z$, and $y-z$), corresponding to a half-flux Hofstadter model in each plane (and a two-site magnetic unit cell); or equivalently a cubic lattice of spin-1/2's with engineered spin-dependent tunneling. In either case, it is a 3D time-reversal symmetric tight-binding model on a cubic lattice; the two sites in the \emph{magnetic} unit cell generate the pseudo-spin degree of freedom in the spin model (see Fig.~\ref{fig:setup}).

The dispersion relation of this tight-binding Hamiltonian is (see SI~\ref{SI:WeylHam}): $\mathcal{H}(\bm{k})=\varepsilon_0  +\bm{h}(\bm{k})\cdot \bm{\sigma}$, with $\bm{h}(\bm{k})/2t_0= \cos(k_x a)\hat{x}-\sin(k_y a)\hat{y}+\cos(k_z a)\hat{z}$. The eigen-energies are thus:
$E_\pm(\bm{k})=\varepsilon_0  \pm 2t_0 \sqrt{\cos^2(k_z a)+ \cos^2(k_x a)+\sin^2(k_y a)}$. The four Weyl points are located at $\bm{k}a=\pi/2(\pm 1,0,\pm 1)$, with energy $E=\varepsilon_0$ and chiral charges $\chi=-\,\text{sgn}(k_x k_z)$. 

The crucial technique required to realize this Weyl model in a circuit is the ability to generate a synthetic magnetic flux by controlling  tunneling phase, which we achieve by capacitively coupling each end of each on-site inductor its neighbors; swapping the connections generates a $\pi$ phase shift in the tunneling amplitude~\cite{ningyuan2015time}.

We assemble an $8\times 8\times 8$ lattice of 2-site unit cells by stacking printed circuit boards (PCBs) with connectivity in $x$, $y$, and $z$ directions shown in Fig.~\ref{fig:setup}b. The lattice translation vectors are $\hat{u}\equiv\frac{1}{\sqrt{2}}(\hat{x}+\hat{y})$, and $\hat{v}\equiv\frac{1}{\sqrt{2}}(\hat{x}-\hat{y})$, and $\hat{z}$, parallel to physical edges of the circuit boards. For the chosen component values (see SI~\ref{SI:WeylCircuitComponents}), the predicted Weyl point frequency is $f_{\rm Weyl}=290$ kHz, and the band-structure spans the frequency range $230-450$ kHz.

\begin{figure*}
\centering
\includegraphics[width=\textwidth]{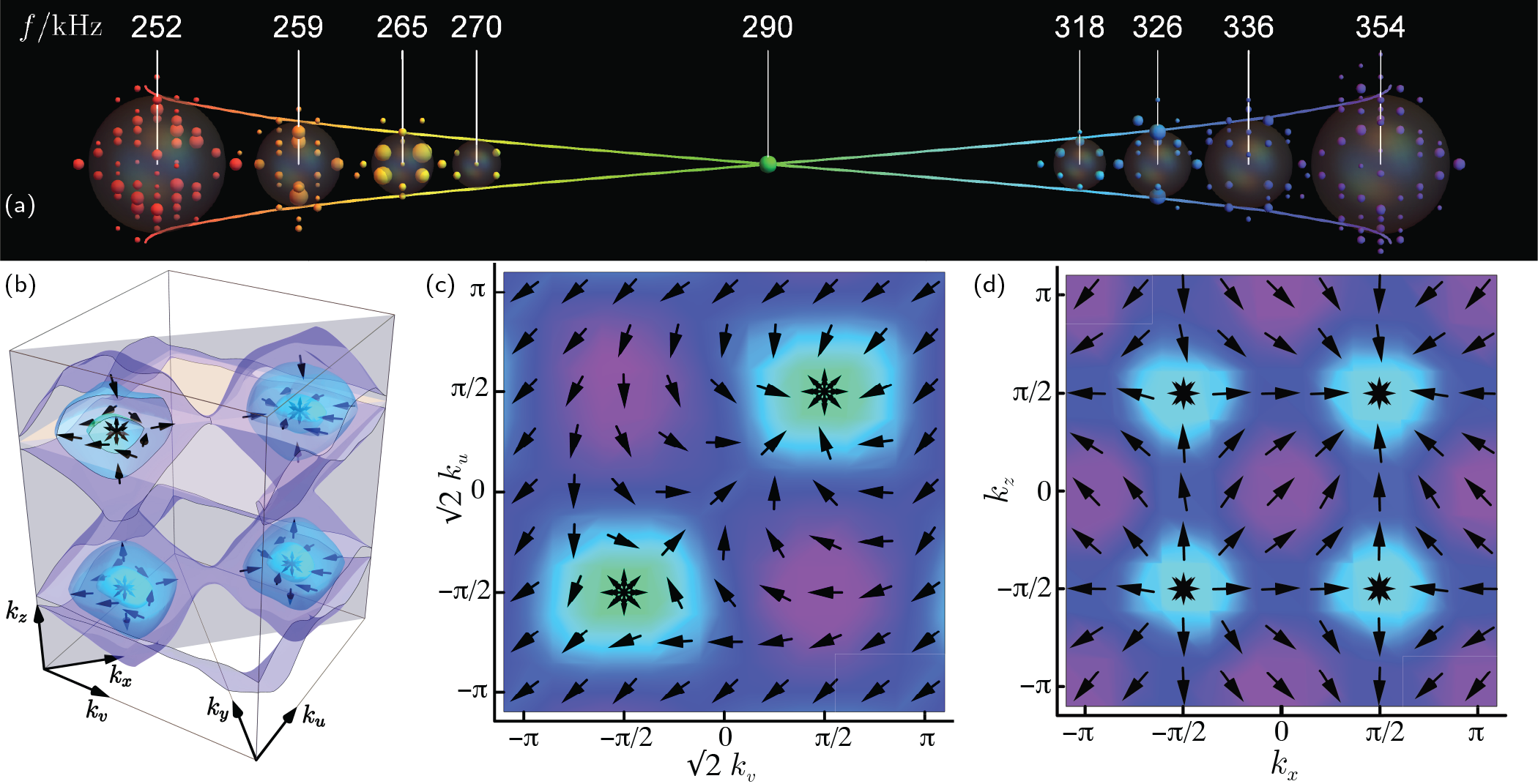}
\caption{\label{fig:bandstructure} \textbf{Measured Band Structure of the Weyl Circuit}. In a configuration with periodic boundary conditions \emph{physically imposed} along all three axes, it is possible to fully reconstruct the bulk band-structure by exciting a single site and measuring the complex response at all sites as a function of frequency. A 3D spatial Fourier transform of the resulting response yields the dispersion shown in \textbf{(a)} in the vicinity of the Weyl point at $\bm{k}=(\pi/2,0,\pi/2)$. Away from the frequency of the Weyl point (red to purple) the system responds primarily on near-spherical constant-energy surfaces in k-space. At the energy of the Weyl point (green), the response is localized \emph{exclusively} to the momentum of the Weyl point.  The equi-energy surfaces are plotted over the full Brillouin zone in \textbf{(b)}, for the band above the Weyl frequency (color-scale same as \textbf{(a)}). The equi-energy surfaces are aspheric at frequencies far from the Weyl points, a generic result arising from the eventual merger of the Weyl cones into a single band. Also plotted (arrows) is the reconstructed spin-texture of the upper band Bloch functions near the Weyl points; at $\bm{k}=(\pi/2,0,\pi/2)$ the spin points everywhere inwardly - a radial hedgehog defect, while the other three Weyl points exhibit hyperbolic hedgehog defects, from which we can deduce the chirality of the Weyl points: $\chi=-\,\text{sgn}(k_x k_z)$. In \textbf{(c)} we plot a slice of the measured dispersion in $k_u-k_v$ plane at the $k_z=\pi/2$ of two Weyl points. In \textbf{(d)} we plot a slice in the $k_x-k_z$  plane at the $k_y=0$  of all Weyl points. \jcmt{With color from green to red labeling the frequency from the Weyl frequency to above also used in \textbf{(a)} and \textbf{(b)},} Also shown, as arrows, are the projections of all the spin-texture into the planes.}
\end{figure*}

\section{Probing Bulk Topology}
To probe the bulk band structure without surface-state contamination, we harness the unique control of our circuit realization~\cite{ningyuan2015time} to impose periodic boundary conditions on all surfaces (See SI~\ref{fig:bandstructure}), thus realizing a finite system which is \emph{all bulk}.

We experimentally extract the band structure and spin texture of the Weyl circuit by measuring the frequency-dependent response of the circuit as a function of spatial offset using a custom-built 3D scanner (see SI~\ref{SI:Device}). Because we recover the full magnitude and phase of the response, we are able to reconstruct (via a Fourier transform) the momentum-dependent response of the system. Plotting this response vs $\vec{k}$ yields the lattice-photon dispersion shown in Fig.~\ref{fig:bandstructure}a, plotted in the vicinity of the Weyl point at $\bm{k}=(-\pi/2,0,-\pi/2)$; at frequencies just above (or below) the Weyl frequency $f_{\rm Weyl}$ the equi-energy surface is a near-sphere around the Weyl point-- as the frequency approaches $f_{\rm Weyl}$, the response collapses to into the Weyl point. In Fig.~\ref{fig:bandstructure}b, equi-energy surfaces of the response are plotted over the full Brillouin zone for frequencies above $f_{\rm Weyl}$, revealing four Weyl points whose momenta are, as anticipated $\bm{k}=(\pm\pi/2,0,\pm\pi/2)$. 

Our ability to resolve the complex-valued response on both $A$ and $B$ sub-lattices permits a complete reconstruction of the spin-texture of the Bloch states. The arrows in Fig.~\ref{fig:bandstructure}b reflect the measured spin-structure of the momentum states on the upper energy-surface; Fig.~\ref{fig:bandstructure}c-d are cuts at $k_z=\pi/2$ and $k_y = 0$, which are planes embedding two Weyl points, and all four, respectively. \jcmt{From the spin texture in the vicinity of the Weyl points we can deduce their chiralities: $\chi=\text{sgn}\left(\frac{\partial(h_x, h_y, h_z)}{\partial (k_x, k_y, k_z)}\right)\Big|_{\bm{k}\in \text{Weyl}}=-\,\text{sgn}(k_x k_z)$.} It is apparent that only near the Weyl point at $\bm{k}=(\pi/2,0,\pi/2)$ does the spin points everywhere radially - an inward radial hedgehog defect, while the other three Weyl points exhibit hyperbolic hedgehogs; this is because we have chosen a uniform definition of the spin-Bloch-sphere over the full Brillouin zone - each Weyl point may be converted into a radial hedgehog through a local spin-transformation of even parity\jcmt{ ($\sigma_{x,y,z}\rightarrow \sigma_{x',y',z'}$)}, from which the sign of the chiral charge of the corresponding Weyl point may be observed directly - outward/inward hedgehog for positive/negative chiral charge, respectively.

Note that topological considerations require Weyl points to come in pairs with opposite chiral charge~\cite{nielsen1981no}; when discrete translation is the sole remaining symmetry (inversion- and time-reversal- are broken) the BZ may exhibit a single pair of Weyl points; the minimum is two pairs if time-reversal symmetry is preserved. The first Weyl fermions, observed in TaAs~\cite{lv2015experimental, xu2015discovery}, did not break time-reversal symmetry and because of the complex structure of TaAs, exhibited 12 pairs of Weyl points. Our platform exhibits the minimum four pairs of Weyl points allowed for a time-reversal symmetric system.

\begin{figure*}
\includegraphics[width=\textwidth]{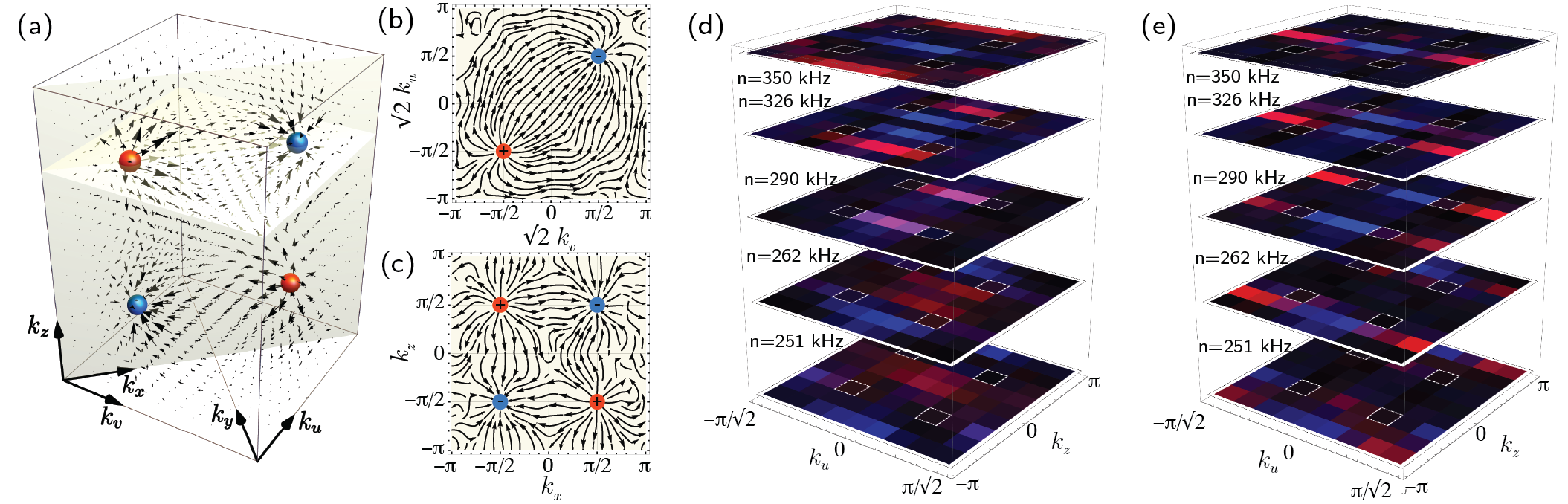}
\caption{\label{fig:curvandsurf}\textbf{Measured Berry Curvature, Berry Flux and Surface States.} \textbf{(a)} shows the measured and interpolated Berry curvature (arrows), and associated Berry flux (blue$\rightarrow$orange is negative$\rightarrow$positive), over the full Brillouin zone, for the lower band. \jcmt{(conventionally, for a 2-band model, the Berry curvature is calculated over the lower band, which is the occupied band for a half-filled Fermionic system)}. Divergence of the Berry curvature is strictly zero except at four topological defects located at the Weyl points, carrying $\pm 1$ chiral charge, with $+1$ corresponding to sources of Berry flux (orange), and $-1$ to sinks (blue), evident from the flow of the Berry curvature. Slices along $k_u-k_v$ and $k_x-k_z$ are shown in \textbf{(b)} and \textbf{(c)} respectively, highlighting the structure of this flow. \textbf{(d)} \& \textbf{(e)} show the measured surface states for even- and odd- numbers of layers in the $v$ direction, respectively. States residing on the top surface are depicted in red, while those on the bottom surface are depicted in blue. It is apparent that, in the $\nu=f_{\rm Weyl}=290 \text{kHz}$ plane, the surface-projection of the Weyl points (outlined in dashed white squares) are connected by lines (``Fermi-arcs'') of surface-states; for an even number of layers, the Fermi-arcs are within the Brillouin zone and overlap each other on both top and bottom surfaces, while for an odd number of layers the Fermi-arc on the top surface (red) passes \emph{around} the Brillouin zone, exploiting its toroidal topology-- a consequence of the gauge difference between even and odd layers (see SI:~\ref{SI:15_layers}).}
\end{figure*}

From the measured space- and spin- resolved Bloch-functions we are able to extract the Berry curvature pseudo-vector of the lower band $\bm{\Omega}(\bm{k})$ via (see SI~\ref{SI:BerryCurvature}): $\bm{\Omega}^{(n)}(\bm{k})=i \mel{\bm{\nabla_{k}} \psi_n(\bm{k})}{\times}{\bm{\nabla_{k}}\psi_n(\bm{k})}$. The curvature is plotted in Fig.~\ref{fig:curvandsurf}a, and it is apparent that while the spin-texture is gauge-dependent, the Berry curvature is \emph{not} -- each Weyl point acts as either a source or sink of Berry curvature, evident from the flow of curvature into/out-of the points. The measured Berry curvature flow from sources to the sinks is shown in Fig.~\ref{fig:curvandsurf}b-c, where 2D slices of the full 3D Brillouin zone are displayed. The overlaid density plot depicts the measured chiral charge density $\rho_\chi\equiv\frac{1}{2\pi}\nabla_k\cdot\bm{\Omega}$, exhibiting two (maximally localized, delta-function-like) sources (positively charged, orange) and two sinks (negatively charged, blue), located at the four Weyl points. Integrating the Berry-flux over a surface enclosing a single Weyl point yields chiral charges of $\chi=\frac{1}{2\pi}\oiint \bm{\Omega} \cdot \hat{\bm{n}} \mathop{}\!\mathrm{d} S=\iiint \rho_{\chi} \mathrm{d}V = +1, -1, +1, -1$, in agreement with theory (see SI~\ref{SI:BerryCurvature}).

\section{Probing Surface States}
The chiral charge of the Weyl nodes is also reflected in the structure of the surface states. This is most easily understood~\cite{hasan2017discovery} by considering a simple material with only one pair of Weyl nodes, and examining the states on an infinite cylindrical surface whose axis $\hat{n}$ is aligned with the vector connecting the (oppositely charged) Weyl points. The resulting 2D band structure, when computed as a function of the axial momentum $k_n$, will exhibit chiral edge states when the Chern number at fixed $k_n$ is non-zero (see SI~\ref{SI:SurfaceBulk}). The momentum-structure of these surface states is model-dependent, but the surface channel must terminate at the projections of the Weyl points onto the surface, yielding ``Fermi arcs'' at $f_{\rm Weyl}$ connecting the Weyl points; additional Weyl points (as in our experiment) result in more arcs. More broadly, Fermi arcs only present on particular surfaces (see SI~\ref{SI:SurfaceBulk}).

We experimentally probe the surface physics by removing the periodic boundaries on the $\hat{v}$ axis (see SI~\ref{SI:RealizeBoundary}) and repeating our site-resolved measurements on the faces of the resulting three-cylinder. The measured surface band-structure is shown in Fig.~\ref{fig:curvandsurf}d-e; the Fermi arcs manifest as lines in $k_u-k_z$ space, linking the surface-projections of the pairs of Weyl points at a frequency of $290$ kHz. When the meta-material has an even number of layers, the arcs on the top- and bottom surfaces follow the same path through the BZ. When the meta-material has an odd number of layers, the Fermi arcs on the top surface connect the Weyl points through the BZ, and around the edge of the (toroidal) BZ on the bottom surface, as anticipated theoretically~\cite{Hosur2012}.

In conclusion, we have realized the first circuit supporting Weyl particles as excitations and explored its properties: We find two pairs of Weyl points, and by spin-resolved spectroscopy directly map out the Berry curvature over the full Brillouin zone, thereby ascertaining that the Weyl points in each pair have opposite quantized chiral charge. Further evidence of the chiral charge of the Weyl points comes from the direct detection of Fermi-arc surface states connecting their surface-projections. The addition of ferrites will enable realization of even more robust T-broken models~\cite{owens2018quarter}, while application of quantum circuit techniques~\cite{anderson2016engineering} will enable qubit-mediated interactions~\cite{wallraff2004strong}, enabling study of the interplay of topology and many-body physics~\cite{roy2017interacting,Chan2017, buividovich2014lattice}.

\section{Methods}
\label{SI:Methods}
The lattice is composed of 8 stacked printed circuit boards (PCBs), each containing two interleaved $8\times 8$ sub-lattices of inductors, capacitively coupled on both ends to their nearest neighbors (See  SI~\ref{SI:WeylCircuitComponents} for details). The 8 boards are further capacitively coupled together through additional inter-board headers. Periodic boundary conditions are imposed by connecting the opposite faces of the bulk together using ribbon cables (see SI~\ref{SI:RealizeBoundary} for details).

Each unit cell is composed of $2\times$ $1.02(1)$ mH inductors, and $12\times$ $100(2)$ pF capacitors. A single inductor-capacitor pair has a measured Q-factor of $136(9)$ (in the frequency range from $233(3)$ kHz to $409(3)$ kHz), limited by inductor ohmic loss.

We probe the lattice using an RF network analyzer to excite a single lattice site (inductor) via a magnetically-coupled drive coil, and measure site-by-site using a magnetic pick-up coil that is translated from site-to-site in the lattice using a heavily modified 3D printer.

The circuit platform offers unique benefits including: \emph{the ability to perform site- resolved measurements of (complex) transport coefficients--} a probe coil may be placed inside the bulk to excite and measure the amplitude and phase response at any site; \emph{exquisite control of global topology--} in each dimension one may choose between periodic boundaries and (sharp) open boundaries, the former proposed but unrealized in optical lattices~\cite{lkacki2016quantum}, and the latter only recently achieved using real~\cite{gaunt2012robust,preiss2015strongly,mukherjee2017homogeneous} or synthetic~\cite{stuhl2015visualizing,mancini2015observation,celi2014synthetic} dimensions; more exotic boundary conditions such as Mobi\"us strips~\cite{ningyuan2015time} and Klein bottles are also possible, along with non-euclidean geometries arising from modified connectivity~\cite{biswas2016fractional}.

\section*{Acknowledgements}
The authors would like to thank Michael Levin for fruitful discussions. This work was supported by DOE grant DE-SC0010267 for apparatus construction/data collection and MURI grant FA9550-16-1-0323 for analysis. D.S. acknowledges support from the David and Lucile Packard Foundation. This work was also supported by the University of Chicago Materials Research Science and Engineering Center, which is funded by National Science Foundation under award number DMR-1420709.

\bibliography{sample}

\begin{thebibliography}{57}%
\makeatletter
\providecommand \@ifxundefined [1]{%
 \@ifx{#1\undefined}
}%
\providecommand \@ifnum [1]{%
 \ifnum #1\expandafter \@firstoftwo
 \else \expandafter \@secondoftwo
 \fi
}%
\providecommand \@ifx [1]{%
 \ifx #1\expandafter \@firstoftwo
 \else \expandafter \@secondoftwo
 \fi
}%
\providecommand \natexlab [1]{#1}%
\providecommand \enquote  [1]{``#1''}%
\providecommand \bibnamefont  [1]{#1}%
\providecommand \bibfnamefont [1]{#1}%
\providecommand \citenamefont [1]{#1}%
\providecommand \href@noop [0]{\@secondoftwo}%
\providecommand \href [0]{\begingroup \@sanitize@url \@href}%
\providecommand \@href[1]{\@@startlink{#1}\@@href}%
\providecommand \@@href[1]{\endgroup#1\@@endlink}%
\providecommand \@sanitize@url [0]{\catcode `\\12\catcode `\$12\catcode
  `\&12\catcode `\#12\catcode `\^12\catcode `\_12\catcode `\%12\relax}%
\providecommand \@@startlink[1]{}%
\providecommand \@@endlink[0]{}%
\providecommand \url  [0]{\begingroup\@sanitize@url \@url }%
\providecommand \@url [1]{\endgroup\@href {#1}{\urlprefix }}%
\providecommand \urlprefix  [0]{URL }%
\providecommand \Eprint [0]{\href }%
\providecommand \doibase [0]{http://dx.doi.org/}%
\providecommand \selectlanguage [0]{\@gobble}%
\providecommand \bibinfo  [0]{\@secondoftwo}%
\providecommand \bibfield  [0]{\@secondoftwo}%
\providecommand \translation [1]{[#1]}%
\providecommand \BibitemOpen [0]{}%
\providecommand \bibitemStop [0]{}%
\providecommand \bibitemNoStop [0]{.\EOS\space}%
\providecommand \EOS [0]{\spacefactor3000\relax}%
\providecommand \BibitemShut  [1]{\csname bibitem#1\endcsname}%
\let\auto@bib@innerbib\@empty
\bibitem [{\citenamefont {Armitage}\ \emph
  {et~al.}(2018{\natexlab{a}})\citenamefont {Armitage}, \citenamefont {Mele},\
  and\ \citenamefont {Vishwanath}}]{armitage2018weyl}%
  \BibitemOpen
  \bibfield  {author} {\bibinfo {author} {\bibfnamefont {N.}~\bibnamefont
  {Armitage}}, \bibinfo {author} {\bibfnamefont {E.}~\bibnamefont {Mele}}, \
  and\ \bibinfo {author} {\bibfnamefont {A.}~\bibnamefont {Vishwanath}},\
  }\href@noop {} {\bibfield  {journal} {\bibinfo  {journal} {Reviews of Modern
  Physics}\ }\textbf {\bibinfo {volume} {90}},\ \bibinfo {pages} {015001}
  (\bibinfo {year} {2018}{\natexlab{a}})}\BibitemShut {NoStop}%
\bibitem [{\citenamefont {Xu}\ \emph {et~al.}(2015)\citenamefont {Xu},
  \citenamefont {Belopolski}, \citenamefont {Alidoust}, \citenamefont
  {Neupane}, \citenamefont {Bian}, \citenamefont {Zhang}, \citenamefont
  {Sankar}, \citenamefont {Chang}, \citenamefont {Yuan}, \citenamefont {Lee}
  \emph {et~al.}}]{xu2015discovery}%
  \BibitemOpen
  \bibfield  {author} {\bibinfo {author} {\bibfnamefont {S.-Y.}\ \bibnamefont
  {Xu}}, \bibinfo {author} {\bibfnamefont {I.}~\bibnamefont {Belopolski}},
  \bibinfo {author} {\bibfnamefont {N.}~\bibnamefont {Alidoust}}, \bibinfo
  {author} {\bibfnamefont {M.}~\bibnamefont {Neupane}}, \bibinfo {author}
  {\bibfnamefont {G.}~\bibnamefont {Bian}}, \bibinfo {author} {\bibfnamefont
  {C.}~\bibnamefont {Zhang}}, \bibinfo {author} {\bibfnamefont
  {R.}~\bibnamefont {Sankar}}, \bibinfo {author} {\bibfnamefont
  {G.}~\bibnamefont {Chang}}, \bibinfo {author} {\bibfnamefont
  {Z.}~\bibnamefont {Yuan}}, \bibinfo {author} {\bibfnamefont {C.-C.}\
  \bibnamefont {Lee}},  \emph {et~al.},\ }\href@noop {} {\bibfield  {journal}
  {\bibinfo  {journal} {Science}\ }\textbf {\bibinfo {volume} {349}},\ \bibinfo
  {pages} {613} (\bibinfo {year} {2015})}\BibitemShut {NoStop}%
\bibitem [{\citenamefont {Lv}\ \emph {et~al.}(2015)\citenamefont {Lv},
  \citenamefont {Weng}, \citenamefont {Fu}, \citenamefont {Wang}, \citenamefont
  {Miao}, \citenamefont {Ma}, \citenamefont {Richard}, \citenamefont {Huang},
  \citenamefont {Zhao}, \citenamefont {Chen} \emph
  {et~al.}}]{lv2015experimental}%
  \BibitemOpen
  \bibfield  {author} {\bibinfo {author} {\bibfnamefont {B.}~\bibnamefont
  {Lv}}, \bibinfo {author} {\bibfnamefont {H.}~\bibnamefont {Weng}}, \bibinfo
  {author} {\bibfnamefont {B.}~\bibnamefont {Fu}}, \bibinfo {author}
  {\bibfnamefont {X.}~\bibnamefont {Wang}}, \bibinfo {author} {\bibfnamefont
  {H.}~\bibnamefont {Miao}}, \bibinfo {author} {\bibfnamefont {J.}~\bibnamefont
  {Ma}}, \bibinfo {author} {\bibfnamefont {P.}~\bibnamefont {Richard}},
  \bibinfo {author} {\bibfnamefont {X.}~\bibnamefont {Huang}}, \bibinfo
  {author} {\bibfnamefont {L.}~\bibnamefont {Zhao}}, \bibinfo {author}
  {\bibfnamefont {G.}~\bibnamefont {Chen}},  \emph {et~al.},\ }\href@noop {}
  {\bibfield  {journal} {\bibinfo  {journal} {Physical Review X}\ }\textbf
  {\bibinfo {volume} {5}},\ \bibinfo {pages} {031013} (\bibinfo {year}
  {2015})}\BibitemShut {NoStop}%
\bibitem [{\citenamefont {Hasan}\ \emph {et~al.}(2017)\citenamefont {Hasan},
  \citenamefont {Xu}, \citenamefont {Belopolski},\ and\ \citenamefont
  {Huang}}]{hasan2017discovery}%
  \BibitemOpen
  \bibfield  {author} {\bibinfo {author} {\bibfnamefont {M.~Z.}\ \bibnamefont
  {Hasan}}, \bibinfo {author} {\bibfnamefont {S.-Y.}\ \bibnamefont {Xu}},
  \bibinfo {author} {\bibfnamefont {I.}~\bibnamefont {Belopolski}}, \ and\
  \bibinfo {author} {\bibfnamefont {S.-M.}\ \bibnamefont {Huang}},\ }\href@noop
  {} {\bibfield  {journal} {\bibinfo  {journal} {Annual Review of Condensed
  Matter Physics}\ }\textbf {\bibinfo {volume} {8}},\ \bibinfo {pages} {289}
  (\bibinfo {year} {2017})}\BibitemShut {NoStop}%
\bibitem [{\citenamefont {Lu}\ \emph {et~al.}(2015)\citenamefont {Lu},
  \citenamefont {Wang}, \citenamefont {Ye}, \citenamefont {Ran}, \citenamefont
  {Fu}, \citenamefont {Joannopoulos},\ and\ \citenamefont
  {Solja{\v{c}}i{\'c}}}]{lu2015experimental}%
  \BibitemOpen
  \bibfield  {author} {\bibinfo {author} {\bibfnamefont {L.}~\bibnamefont
  {Lu}}, \bibinfo {author} {\bibfnamefont {Z.}~\bibnamefont {Wang}}, \bibinfo
  {author} {\bibfnamefont {D.}~\bibnamefont {Ye}}, \bibinfo {author}
  {\bibfnamefont {L.}~\bibnamefont {Ran}}, \bibinfo {author} {\bibfnamefont
  {L.}~\bibnamefont {Fu}}, \bibinfo {author} {\bibfnamefont {J.~D.}\
  \bibnamefont {Joannopoulos}}, \ and\ \bibinfo {author} {\bibfnamefont
  {M.}~\bibnamefont {Solja{\v{c}}i{\'c}}},\ }\href@noop {} {\bibfield
  {journal} {\bibinfo  {journal} {Science}\ }\textbf {\bibinfo {volume}
  {349}},\ \bibinfo {pages} {622} (\bibinfo {year} {2015})}\BibitemShut
  {NoStop}%
\bibitem [{\citenamefont {Noh}\ \emph {et~al.}(2017)\citenamefont {Noh},
  \citenamefont {Huang}, \citenamefont {Leykam}, \citenamefont {Chong},
  \citenamefont {Chen},\ and\ \citenamefont {Rechtsman}}]{noh2017experimental}%
  \BibitemOpen
  \bibfield  {author} {\bibinfo {author} {\bibfnamefont {J.}~\bibnamefont
  {Noh}}, \bibinfo {author} {\bibfnamefont {S.}~\bibnamefont {Huang}}, \bibinfo
  {author} {\bibfnamefont {D.}~\bibnamefont {Leykam}}, \bibinfo {author}
  {\bibfnamefont {Y.}~\bibnamefont {Chong}}, \bibinfo {author} {\bibfnamefont
  {K.~P.}\ \bibnamefont {Chen}}, \ and\ \bibinfo {author} {\bibfnamefont
  {M.~C.}\ \bibnamefont {Rechtsman}},\ }\href@noop {} {\bibfield  {journal}
  {\bibinfo  {journal} {Nature Physics}\ }\textbf {\bibinfo {volume} {13}},\
  \bibinfo {pages} {611} (\bibinfo {year} {2017})}\BibitemShut {NoStop}%
\bibitem [{\citenamefont {Dub{\v{c}}ek}\ \emph {et~al.}(2015)\citenamefont
  {Dub{\v{c}}ek}, \citenamefont {Kennedy}, \citenamefont {Lu}, \citenamefont
  {Ketterle}, \citenamefont {Solja{\v{c}}i{\'c}},\ and\ \citenamefont
  {Buljan}}]{dubvcek2015weyl}%
  \BibitemOpen
  \bibfield  {author} {\bibinfo {author} {\bibfnamefont {T.}~\bibnamefont
  {Dub{\v{c}}ek}}, \bibinfo {author} {\bibfnamefont {C.~J.}\ \bibnamefont
  {Kennedy}}, \bibinfo {author} {\bibfnamefont {L.}~\bibnamefont {Lu}},
  \bibinfo {author} {\bibfnamefont {W.}~\bibnamefont {Ketterle}}, \bibinfo
  {author} {\bibfnamefont {M.}~\bibnamefont {Solja{\v{c}}i{\'c}}}, \ and\
  \bibinfo {author} {\bibfnamefont {H.}~\bibnamefont {Buljan}},\ }\href@noop {}
  {\bibfield  {journal} {\bibinfo  {journal} {Physical review letters}\
  }\textbf {\bibinfo {volume} {114}},\ \bibinfo {pages} {225301} (\bibinfo
  {year} {2015})}\BibitemShut {NoStop}%
\bibitem [{\citenamefont {Ningyuan}\ \emph {et~al.}(2015)\citenamefont
  {Ningyuan}, \citenamefont {Owens}, \citenamefont {Sommer}, \citenamefont
  {Schuster},\ and\ \citenamefont {Simon}}]{ningyuan2015time}%
  \BibitemOpen
  \bibfield  {author} {\bibinfo {author} {\bibfnamefont {J.}~\bibnamefont
  {Ningyuan}}, \bibinfo {author} {\bibfnamefont {C.}~\bibnamefont {Owens}},
  \bibinfo {author} {\bibfnamefont {A.}~\bibnamefont {Sommer}}, \bibinfo
  {author} {\bibfnamefont {D.}~\bibnamefont {Schuster}}, \ and\ \bibinfo
  {author} {\bibfnamefont {J.}~\bibnamefont {Simon}},\ }\href@noop {}
  {\bibfield  {journal} {\bibinfo  {journal} {Physical Review X}\ }\textbf
  {\bibinfo {volume} {5}},\ \bibinfo {pages} {021031} (\bibinfo {year}
  {2015})}\BibitemShut {NoStop}%
\bibitem [{\citenamefont {Wei}\ \emph {et~al.}(2012)\citenamefont {Wei},
  \citenamefont {Chao},\ and\ \citenamefont {Aji}}]{wei2012excitonic}%
  \BibitemOpen
  \bibfield  {author} {\bibinfo {author} {\bibfnamefont {H.}~\bibnamefont
  {Wei}}, \bibinfo {author} {\bibfnamefont {S.-P.}\ \bibnamefont {Chao}}, \
  and\ \bibinfo {author} {\bibfnamefont {V.}~\bibnamefont {Aji}},\ }\href@noop
  {} {\bibfield  {journal} {\bibinfo  {journal} {Physical review letters}\
  }\textbf {\bibinfo {volume} {109}},\ \bibinfo {pages} {196403} (\bibinfo
  {year} {2012})}\BibitemShut {NoStop}%
\bibitem [{\citenamefont {Wallraff}\ \emph {et~al.}(2004)\citenamefont
  {Wallraff}, \citenamefont {Schuster}, \citenamefont {Blais}, \citenamefont
  {Frunzio}, \citenamefont {Huang}, \citenamefont {Majer}, \citenamefont
  {Kumar}, \citenamefont {Girvin},\ and\ \citenamefont
  {Schoelkopf}}]{wallraff2004strong}%
  \BibitemOpen
  \bibfield  {author} {\bibinfo {author} {\bibfnamefont {A.}~\bibnamefont
  {Wallraff}}, \bibinfo {author} {\bibfnamefont {D.~I.}\ \bibnamefont
  {Schuster}}, \bibinfo {author} {\bibfnamefont {A.}~\bibnamefont {Blais}},
  \bibinfo {author} {\bibfnamefont {L.}~\bibnamefont {Frunzio}}, \bibinfo
  {author} {\bibfnamefont {R.-S.}\ \bibnamefont {Huang}}, \bibinfo {author}
  {\bibfnamefont {J.}~\bibnamefont {Majer}}, \bibinfo {author} {\bibfnamefont
  {S.}~\bibnamefont {Kumar}}, \bibinfo {author} {\bibfnamefont {S.~M.}\
  \bibnamefont {Girvin}}, \ and\ \bibinfo {author} {\bibfnamefont {R.~J.}\
  \bibnamefont {Schoelkopf}},\ }\href@noop {} {\bibfield  {journal} {\bibinfo
  {journal} {Nature}\ }\textbf {\bibinfo {volume} {431}},\ \bibinfo {pages}
  {162} (\bibinfo {year} {2004})}\BibitemShut {NoStop}%
\bibitem [{\citenamefont {Weststr{\"o}m}\ and\ \citenamefont
  {Ojanen}(2017)}]{weststrom2017designer}%
  \BibitemOpen
  \bibfield  {author} {\bibinfo {author} {\bibfnamefont {A.}~\bibnamefont
  {Weststr{\"o}m}}\ and\ \bibinfo {author} {\bibfnamefont {T.}~\bibnamefont
  {Ojanen}},\ }\href@noop {} {\bibfield  {journal} {\bibinfo  {journal}
  {Physical Review X}\ }\textbf {\bibinfo {volume} {7}},\ \bibinfo {pages}
  {041026} (\bibinfo {year} {2017})}\BibitemShut {NoStop}%
\bibitem [{\citenamefont {Bloch}\ \emph {et~al.}(2008)\citenamefont {Bloch},
  \citenamefont {Dalibard},\ and\ \citenamefont {Zwerger}}]{bloch2008many}%
  \BibitemOpen
  \bibfield  {author} {\bibinfo {author} {\bibfnamefont {I.}~\bibnamefont
  {Bloch}}, \bibinfo {author} {\bibfnamefont {J.}~\bibnamefont {Dalibard}}, \
  and\ \bibinfo {author} {\bibfnamefont {W.}~\bibnamefont {Zwerger}},\
  }\href@noop {} {\bibfield  {journal} {\bibinfo  {journal} {Reviews of modern
  physics}\ }\textbf {\bibinfo {volume} {80}},\ \bibinfo {pages} {885}
  (\bibinfo {year} {2008})}\BibitemShut {NoStop}%
\bibitem [{\citenamefont {Carusotto}\ and\ \citenamefont
  {Ciuti}(2013)}]{carusotto2013quantum}%
  \BibitemOpen
  \bibfield  {author} {\bibinfo {author} {\bibfnamefont {I.}~\bibnamefont
  {Carusotto}}\ and\ \bibinfo {author} {\bibfnamefont {C.}~\bibnamefont
  {Ciuti}},\ }\href@noop {} {\bibfield  {journal} {\bibinfo  {journal} {Reviews
  of Modern Physics}\ }\textbf {\bibinfo {volume} {85}},\ \bibinfo {pages}
  {299} (\bibinfo {year} {2013})}\BibitemShut {NoStop}%
\bibitem [{\citenamefont {Goldman}\ \emph {et~al.}(2014)\citenamefont
  {Goldman}, \citenamefont {Juzeli{\=u}nas}, \citenamefont {{\"O}hberg},\ and\
  \citenamefont {Spielman}}]{goldman2014light}%
  \BibitemOpen
  \bibfield  {author} {\bibinfo {author} {\bibfnamefont {N.}~\bibnamefont
  {Goldman}}, \bibinfo {author} {\bibfnamefont {G.}~\bibnamefont
  {Juzeli{\=u}nas}}, \bibinfo {author} {\bibfnamefont {P.}~\bibnamefont
  {{\"O}hberg}}, \ and\ \bibinfo {author} {\bibfnamefont {I.~B.}\ \bibnamefont
  {Spielman}},\ }\href@noop {} {\bibfield  {journal} {\bibinfo  {journal}
  {Reports on Progress in Physics}\ }\textbf {\bibinfo {volume} {77}},\
  \bibinfo {pages} {126401} (\bibinfo {year} {2014})}\BibitemShut {NoStop}%
\bibitem [{\citenamefont {Hafezi}\ \emph {et~al.}(2011)\citenamefont {Hafezi},
  \citenamefont {Demler}, \citenamefont {Lukin},\ and\ \citenamefont
  {Taylor}}]{hafezi2011robust}%
  \BibitemOpen
  \bibfield  {author} {\bibinfo {author} {\bibfnamefont {M.}~\bibnamefont
  {Hafezi}}, \bibinfo {author} {\bibfnamefont {E.~A.}\ \bibnamefont {Demler}},
  \bibinfo {author} {\bibfnamefont {M.~D.}\ \bibnamefont {Lukin}}, \ and\
  \bibinfo {author} {\bibfnamefont {J.~M.}\ \bibnamefont {Taylor}},\
  }\href@noop {} {\bibfield  {journal} {\bibinfo  {journal} {Nature Physics}\
  }\textbf {\bibinfo {volume} {7}},\ \bibinfo {pages} {907} (\bibinfo {year}
  {2011})}\BibitemShut {NoStop}%
\bibitem [{\citenamefont {Mahoney}\ \emph {et~al.}(2017)\citenamefont
  {Mahoney}, \citenamefont {Colless}, \citenamefont {Pauka}, \citenamefont
  {Hornibrook}, \citenamefont {Watson}, \citenamefont {Gardner}, \citenamefont
  {Manfra}, \citenamefont {Doherty},\ and\ \citenamefont
  {Reilly}}]{mahoney2017chip}%
  \BibitemOpen
  \bibfield  {author} {\bibinfo {author} {\bibfnamefont {A.}~\bibnamefont
  {Mahoney}}, \bibinfo {author} {\bibfnamefont {J.}~\bibnamefont {Colless}},
  \bibinfo {author} {\bibfnamefont {S.}~\bibnamefont {Pauka}}, \bibinfo
  {author} {\bibfnamefont {J.}~\bibnamefont {Hornibrook}}, \bibinfo {author}
  {\bibfnamefont {J.}~\bibnamefont {Watson}}, \bibinfo {author} {\bibfnamefont
  {G.}~\bibnamefont {Gardner}}, \bibinfo {author} {\bibfnamefont
  {M.}~\bibnamefont {Manfra}}, \bibinfo {author} {\bibfnamefont
  {A.}~\bibnamefont {Doherty}}, \ and\ \bibinfo {author} {\bibfnamefont
  {D.}~\bibnamefont {Reilly}},\ }\href@noop {} {\bibfield  {journal} {\bibinfo
  {journal} {Physical Review X}\ }\textbf {\bibinfo {volume} {7}},\ \bibinfo
  {pages} {011007} (\bibinfo {year} {2017})}\BibitemShut {NoStop}%
\bibitem [{\citenamefont {Tarruell}\ \emph {et~al.}(2012)\citenamefont
  {Tarruell}, \citenamefont {Greif}, \citenamefont {Uehlinger}, \citenamefont
  {Jotzu},\ and\ \citenamefont {Esslinger}}]{tarruell2012creating}%
  \BibitemOpen
  \bibfield  {author} {\bibinfo {author} {\bibfnamefont {L.}~\bibnamefont
  {Tarruell}}, \bibinfo {author} {\bibfnamefont {D.}~\bibnamefont {Greif}},
  \bibinfo {author} {\bibfnamefont {T.}~\bibnamefont {Uehlinger}}, \bibinfo
  {author} {\bibfnamefont {G.}~\bibnamefont {Jotzu}}, \ and\ \bibinfo {author}
  {\bibfnamefont {T.}~\bibnamefont {Esslinger}},\ }\href@noop {} {\bibfield
  {journal} {\bibinfo  {journal} {Nature}\ }\textbf {\bibinfo {volume} {483}},\
  \bibinfo {pages} {302} (\bibinfo {year} {2012})}\BibitemShut {NoStop}%
\bibitem [{\citenamefont {Gomes}\ \emph {et~al.}(2012)\citenamefont {Gomes},
  \citenamefont {Mar}, \citenamefont {Ko}, \citenamefont {Guinea},\ and\
  \citenamefont {Manoharan}}]{gomes2012designer}%
  \BibitemOpen
  \bibfield  {author} {\bibinfo {author} {\bibfnamefont {K.~K.}\ \bibnamefont
  {Gomes}}, \bibinfo {author} {\bibfnamefont {W.}~\bibnamefont {Mar}}, \bibinfo
  {author} {\bibfnamefont {W.}~\bibnamefont {Ko}}, \bibinfo {author}
  {\bibfnamefont {F.}~\bibnamefont {Guinea}}, \ and\ \bibinfo {author}
  {\bibfnamefont {H.~C.}\ \bibnamefont {Manoharan}},\ }\href@noop {} {\bibfield
   {journal} {\bibinfo  {journal} {Nature}\ }\textbf {\bibinfo {volume}
  {483}},\ \bibinfo {pages} {306} (\bibinfo {year} {2012})}\BibitemShut
  {NoStop}%
\bibitem [{\citenamefont {Jotzu}\ \emph {et~al.}(2014)\citenamefont {Jotzu},
  \citenamefont {Messer}, \citenamefont {Desbuquois}, \citenamefont {Lebrat},
  \citenamefont {Uehlinger}, \citenamefont {Greif},\ and\ \citenamefont
  {Esslinger}}]{jotzu2014experimental}%
  \BibitemOpen
  \bibfield  {author} {\bibinfo {author} {\bibfnamefont {G.}~\bibnamefont
  {Jotzu}}, \bibinfo {author} {\bibfnamefont {M.}~\bibnamefont {Messer}},
  \bibinfo {author} {\bibfnamefont {R.}~\bibnamefont {Desbuquois}}, \bibinfo
  {author} {\bibfnamefont {M.}~\bibnamefont {Lebrat}}, \bibinfo {author}
  {\bibfnamefont {T.}~\bibnamefont {Uehlinger}}, \bibinfo {author}
  {\bibfnamefont {D.}~\bibnamefont {Greif}}, \ and\ \bibinfo {author}
  {\bibfnamefont {T.}~\bibnamefont {Esslinger}},\ }\href@noop {} {\bibfield
  {journal} {\bibinfo  {journal} {Nature}\ }\textbf {\bibinfo {volume} {515}},\
  \bibinfo {pages} {237} (\bibinfo {year} {2014})}\BibitemShut {NoStop}%
\bibitem [{\citenamefont {Lin}\ \emph {et~al.}(2011)\citenamefont {Lin},
  \citenamefont {Jim{\'e}nez-Garc{\'\i}a},\ and\ \citenamefont
  {Spielman}}]{lin2011spin}%
  \BibitemOpen
  \bibfield  {author} {\bibinfo {author} {\bibfnamefont {Y.-J.}\ \bibnamefont
  {Lin}}, \bibinfo {author} {\bibfnamefont {K.}~\bibnamefont
  {Jim{\'e}nez-Garc{\'\i}a}}, \ and\ \bibinfo {author} {\bibfnamefont {I.~B.}\
  \bibnamefont {Spielman}},\ }\href@noop {} {\bibfield  {journal} {\bibinfo
  {journal} {Nature}\ }\textbf {\bibinfo {volume} {471}},\ \bibinfo {pages}
  {83} (\bibinfo {year} {2011})}\BibitemShut {NoStop}%
\bibitem [{\citenamefont {Wang}\ \emph {et~al.}(2018)\citenamefont {Wang},
  \citenamefont {Lu}, \citenamefont {Sun}, \citenamefont {Chen}, \citenamefont
  {Deng},\ and\ \citenamefont {Liu}}]{Wang2018}%
  \BibitemOpen
  \bibfield  {author} {\bibinfo {author} {\bibfnamefont {B.-Z.}\ \bibnamefont
  {Wang}}, \bibinfo {author} {\bibfnamefont {Y.-H.}\ \bibnamefont {Lu}},
  \bibinfo {author} {\bibfnamefont {W.}~\bibnamefont {Sun}}, \bibinfo {author}
  {\bibfnamefont {S.}~\bibnamefont {Chen}}, \bibinfo {author} {\bibfnamefont
  {Y.}~\bibnamefont {Deng}}, \ and\ \bibinfo {author} {\bibfnamefont {X.-J.}\
  \bibnamefont {Liu}},\ }\href@noop {} {\bibfield  {journal} {\bibinfo
  {journal} {Phys. Rev. A}\ }\textbf {\bibinfo {volume} {97}},\ \bibinfo
  {pages} {011605} (\bibinfo {year} {2018})}\BibitemShut {NoStop}%
\bibitem [{\citenamefont {Lin}\ \emph {et~al.}(2009)\citenamefont {Lin},
  \citenamefont {Compton}, \citenamefont {Jimenez-Garcia}, \citenamefont
  {Porto},\ and\ \citenamefont {Spielman}}]{lin2009synthetic}%
  \BibitemOpen
  \bibfield  {author} {\bibinfo {author} {\bibfnamefont {Y.-J.}\ \bibnamefont
  {Lin}}, \bibinfo {author} {\bibfnamefont {R.~L.}\ \bibnamefont {Compton}},
  \bibinfo {author} {\bibfnamefont {K.}~\bibnamefont {Jimenez-Garcia}},
  \bibinfo {author} {\bibfnamefont {J.~V.}\ \bibnamefont {Porto}}, \ and\
  \bibinfo {author} {\bibfnamefont {I.~B.}\ \bibnamefont {Spielman}},\
  }\href@noop {} {\bibfield  {journal} {\bibinfo  {journal} {Nature}\ }\textbf
  {\bibinfo {volume} {462}},\ \bibinfo {pages} {628} (\bibinfo {year}
  {2009})}\BibitemShut {NoStop}%
\bibitem [{\citenamefont {Hafezi}\ \emph {et~al.}(2013)\citenamefont {Hafezi},
  \citenamefont {Mittal}, \citenamefont {Fan}, \citenamefont {Migdall},\ and\
  \citenamefont {Taylor}}]{hafezi2013imaging}%
  \BibitemOpen
  \bibfield  {author} {\bibinfo {author} {\bibfnamefont {M.}~\bibnamefont
  {Hafezi}}, \bibinfo {author} {\bibfnamefont {S.}~\bibnamefont {Mittal}},
  \bibinfo {author} {\bibfnamefont {J.}~\bibnamefont {Fan}}, \bibinfo {author}
  {\bibfnamefont {A.}~\bibnamefont {Migdall}}, \ and\ \bibinfo {author}
  {\bibfnamefont {J.}~\bibnamefont {Taylor}},\ }\href@noop {} {\bibfield
  {journal} {\bibinfo  {journal} {Nature Photonics}\ }\textbf {\bibinfo
  {volume} {7}},\ \bibinfo {pages} {1001} (\bibinfo {year} {2013})}\BibitemShut
  {NoStop}%
\bibitem [{\citenamefont {Wang}\ \emph {et~al.}(2009)\citenamefont {Wang},
  \citenamefont {Chong}, \citenamefont {Joannopoulos},\ and\ \citenamefont
  {Solja{\v{c}}i{\'c}}}]{wang2009observation}%
  \BibitemOpen
  \bibfield  {author} {\bibinfo {author} {\bibfnamefont {Z.}~\bibnamefont
  {Wang}}, \bibinfo {author} {\bibfnamefont {Y.}~\bibnamefont {Chong}},
  \bibinfo {author} {\bibfnamefont {J.~D.}\ \bibnamefont {Joannopoulos}}, \
  and\ \bibinfo {author} {\bibfnamefont {M.}~\bibnamefont
  {Solja{\v{c}}i{\'c}}},\ }\href@noop {} {\bibfield  {journal} {\bibinfo
  {journal} {Nature}\ }\textbf {\bibinfo {volume} {461}},\ \bibinfo {pages}
  {772} (\bibinfo {year} {2009})}\BibitemShut {NoStop}%
\bibitem [{\citenamefont {Rechtsman}\ \emph {et~al.}(2013)\citenamefont
  {Rechtsman}, \citenamefont {Zeuner}, \citenamefont {Plotnik}, \citenamefont
  {Lumer}, \citenamefont {Podolsky}, \citenamefont {Dreisow}, \citenamefont
  {Nolte}, \citenamefont {Segev},\ and\ \citenamefont
  {Szameit}}]{rechtsman2013photonic}%
  \BibitemOpen
  \bibfield  {author} {\bibinfo {author} {\bibfnamefont {M.~C.}\ \bibnamefont
  {Rechtsman}}, \bibinfo {author} {\bibfnamefont {J.~M.}\ \bibnamefont
  {Zeuner}}, \bibinfo {author} {\bibfnamefont {Y.}~\bibnamefont {Plotnik}},
  \bibinfo {author} {\bibfnamefont {Y.}~\bibnamefont {Lumer}}, \bibinfo
  {author} {\bibfnamefont {D.}~\bibnamefont {Podolsky}}, \bibinfo {author}
  {\bibfnamefont {F.}~\bibnamefont {Dreisow}}, \bibinfo {author} {\bibfnamefont
  {S.}~\bibnamefont {Nolte}}, \bibinfo {author} {\bibfnamefont
  {M.}~\bibnamefont {Segev}}, \ and\ \bibinfo {author} {\bibfnamefont
  {A.}~\bibnamefont {Szameit}},\ }\href@noop {} {\bibfield  {journal} {\bibinfo
   {journal} {Nature}\ }\textbf {\bibinfo {volume} {496}},\ \bibinfo {pages}
  {196} (\bibinfo {year} {2013})}\BibitemShut {NoStop}%
\bibitem [{\citenamefont {Owens}\ \emph {et~al.}(2018)\citenamefont {Owens},
  \citenamefont {LaChapelle}, \citenamefont {Saxberg}, \citenamefont
  {Anderson}, \citenamefont {Ma}, \citenamefont {Simon},\ and\ \citenamefont
  {Schuster}}]{owens2018quarter}%
  \BibitemOpen
  \bibfield  {author} {\bibinfo {author} {\bibfnamefont {C.}~\bibnamefont
  {Owens}}, \bibinfo {author} {\bibfnamefont {A.}~\bibnamefont {LaChapelle}},
  \bibinfo {author} {\bibfnamefont {B.}~\bibnamefont {Saxberg}}, \bibinfo
  {author} {\bibfnamefont {B.~M.}\ \bibnamefont {Anderson}}, \bibinfo {author}
  {\bibfnamefont {R.}~\bibnamefont {Ma}}, \bibinfo {author} {\bibfnamefont
  {J.}~\bibnamefont {Simon}}, \ and\ \bibinfo {author} {\bibfnamefont {D.~I.}\
  \bibnamefont {Schuster}},\ }\href@noop {} {\bibfield  {journal} {\bibinfo
  {journal} {Physical Review A}\ }\textbf {\bibinfo {volume} {97}},\ \bibinfo
  {pages} {013818} (\bibinfo {year} {2018})}\BibitemShut {NoStop}%
\bibitem [{\citenamefont {Schine}\ \emph {et~al.}(2016)\citenamefont {Schine},
  \citenamefont {Ryou}, \citenamefont {Gromov}, \citenamefont {Sommer},\ and\
  \citenamefont {Simon}}]{schine2016synthetic}%
  \BibitemOpen
  \bibfield  {author} {\bibinfo {author} {\bibfnamefont {N.}~\bibnamefont
  {Schine}}, \bibinfo {author} {\bibfnamefont {A.}~\bibnamefont {Ryou}},
  \bibinfo {author} {\bibfnamefont {A.}~\bibnamefont {Gromov}}, \bibinfo
  {author} {\bibfnamefont {A.}~\bibnamefont {Sommer}}, \ and\ \bibinfo {author}
  {\bibfnamefont {J.}~\bibnamefont {Simon}},\ }\href@noop {} {\bibfield
  {journal} {\bibinfo  {journal} {Nature}\ }\textbf {\bibinfo {volume} {534}},\
  \bibinfo {pages} {671} (\bibinfo {year} {2016})}\BibitemShut {NoStop}%
\bibitem [{\citenamefont {S{\"u}sstrunk}\ and\ \citenamefont
  {Huber}(2015)}]{susstrunk2015observation}%
  \BibitemOpen
  \bibfield  {author} {\bibinfo {author} {\bibfnamefont {R.}~\bibnamefont
  {S{\"u}sstrunk}}\ and\ \bibinfo {author} {\bibfnamefont {S.~D.}\ \bibnamefont
  {Huber}},\ }\href@noop {} {\bibfield  {journal} {\bibinfo  {journal}
  {Science}\ }\textbf {\bibinfo {volume} {349}},\ \bibinfo {pages} {47}
  (\bibinfo {year} {2015})}\BibitemShut {NoStop}%
\bibitem [{\citenamefont {Nash}\ \emph {et~al.}(2015)\citenamefont {Nash},
  \citenamefont {Kleckner}, \citenamefont {Read}, \citenamefont {Vitelli},
  \citenamefont {Turner},\ and\ \citenamefont {Irvine}}]{nash2015topological}%
  \BibitemOpen
  \bibfield  {author} {\bibinfo {author} {\bibfnamefont {L.~M.}\ \bibnamefont
  {Nash}}, \bibinfo {author} {\bibfnamefont {D.}~\bibnamefont {Kleckner}},
  \bibinfo {author} {\bibfnamefont {A.}~\bibnamefont {Read}}, \bibinfo {author}
  {\bibfnamefont {V.}~\bibnamefont {Vitelli}}, \bibinfo {author} {\bibfnamefont
  {A.~M.}\ \bibnamefont {Turner}}, \ and\ \bibinfo {author} {\bibfnamefont
  {W.~T.}\ \bibnamefont {Irvine}},\ }\href@noop {} {\bibfield  {journal}
  {\bibinfo  {journal} {Proceedings of the National Academy of Sciences}\
  }\textbf {\bibinfo {volume} {112}},\ \bibinfo {pages} {14495} (\bibinfo
  {year} {2015})}\BibitemShut {NoStop}%
\bibitem [{\citenamefont {Tai}\ \emph {et~al.}(2017)\citenamefont {Tai},
  \citenamefont {Lukin}, \citenamefont {Rispoli}, \citenamefont {Schittko},
  \citenamefont {Menke}, \citenamefont {Borgnia}, \citenamefont {Preiss},
  \citenamefont {Grusdt}, \citenamefont {Kaufman},\ and\ \citenamefont
  {Greiner}}]{tai2017microscopy}%
  \BibitemOpen
  \bibfield  {author} {\bibinfo {author} {\bibfnamefont {M.~E.}\ \bibnamefont
  {Tai}}, \bibinfo {author} {\bibfnamefont {A.}~\bibnamefont {Lukin}}, \bibinfo
  {author} {\bibfnamefont {M.}~\bibnamefont {Rispoli}}, \bibinfo {author}
  {\bibfnamefont {R.}~\bibnamefont {Schittko}}, \bibinfo {author}
  {\bibfnamefont {T.}~\bibnamefont {Menke}}, \bibinfo {author} {\bibfnamefont
  {D.}~\bibnamefont {Borgnia}}, \bibinfo {author} {\bibfnamefont {P.~M.}\
  \bibnamefont {Preiss}}, \bibinfo {author} {\bibfnamefont {F.}~\bibnamefont
  {Grusdt}}, \bibinfo {author} {\bibfnamefont {A.~M.}\ \bibnamefont {Kaufman}},
  \ and\ \bibinfo {author} {\bibfnamefont {M.}~\bibnamefont {Greiner}},\
  }\href@noop {} {\bibfield  {journal} {\bibinfo  {journal} {Nature}\ }\textbf
  {\bibinfo {volume} {546}},\ \bibinfo {pages} {519} (\bibinfo {year}
  {2017})}\BibitemShut {NoStop}%
\bibitem [{\citenamefont {Lewenstein}\ \emph {et~al.}(2007)\citenamefont
  {Lewenstein}, \citenamefont {Sanpera}, \citenamefont {Ahufinger},
  \citenamefont {Damski}, \citenamefont {Sen},\ and\ \citenamefont
  {Sen}}]{lewenstein2007ultracold}%
  \BibitemOpen
  \bibfield  {author} {\bibinfo {author} {\bibfnamefont {M.}~\bibnamefont
  {Lewenstein}}, \bibinfo {author} {\bibfnamefont {A.}~\bibnamefont {Sanpera}},
  \bibinfo {author} {\bibfnamefont {V.}~\bibnamefont {Ahufinger}}, \bibinfo
  {author} {\bibfnamefont {B.}~\bibnamefont {Damski}}, \bibinfo {author}
  {\bibfnamefont {A.}~\bibnamefont {Sen}}, \ and\ \bibinfo {author}
  {\bibfnamefont {U.}~\bibnamefont {Sen}},\ }\href@noop {} {\bibfield
  {journal} {\bibinfo  {journal} {Advances In Physics}\ }\textbf {\bibinfo
  {volume} {56}},\ \bibinfo {pages} {243} (\bibinfo {year} {2007})}\BibitemShut
  {NoStop}%
\bibitem [{\citenamefont {Peyronel}\ \emph {et~al.}(2012)\citenamefont
  {Peyronel}, \citenamefont {Firstenberg}, \citenamefont {Liang}, \citenamefont
  {Hofferberth}, \citenamefont {Gorshkov}, \citenamefont {Pohl}, \citenamefont
  {Lukin},\ and\ \citenamefont {Vuleti{\'c}}}]{peyronel2012quantum}%
  \BibitemOpen
  \bibfield  {author} {\bibinfo {author} {\bibfnamefont {T.}~\bibnamefont
  {Peyronel}}, \bibinfo {author} {\bibfnamefont {O.}~\bibnamefont
  {Firstenberg}}, \bibinfo {author} {\bibfnamefont {Q.-Y.}\ \bibnamefont
  {Liang}}, \bibinfo {author} {\bibfnamefont {S.}~\bibnamefont {Hofferberth}},
  \bibinfo {author} {\bibfnamefont {A.~V.}\ \bibnamefont {Gorshkov}}, \bibinfo
  {author} {\bibfnamefont {T.}~\bibnamefont {Pohl}}, \bibinfo {author}
  {\bibfnamefont {M.~D.}\ \bibnamefont {Lukin}}, \ and\ \bibinfo {author}
  {\bibfnamefont {V.}~\bibnamefont {Vuleti{\'c}}},\ }\href@noop {} {\bibfield
  {journal} {\bibinfo  {journal} {Nature}\ }\textbf {\bibinfo {volume} {488}},\
  \bibinfo {pages} {57} (\bibinfo {year} {2012})}\BibitemShut {NoStop}%
\bibitem [{\citenamefont {Jia}\ \emph {et~al.}(2017)\citenamefont {Jia},
  \citenamefont {Schine}, \citenamefont {Georgakopoulos}, \citenamefont {Ryou},
  \citenamefont {Sommer},\ and\ \citenamefont {Simon}}]{jia2017strongly}%
  \BibitemOpen
  \bibfield  {author} {\bibinfo {author} {\bibfnamefont {N.}~\bibnamefont
  {Jia}}, \bibinfo {author} {\bibfnamefont {N.}~\bibnamefont {Schine}},
  \bibinfo {author} {\bibfnamefont {A.}~\bibnamefont {Georgakopoulos}},
  \bibinfo {author} {\bibfnamefont {A.}~\bibnamefont {Ryou}}, \bibinfo {author}
  {\bibfnamefont {A.}~\bibnamefont {Sommer}}, \ and\ \bibinfo {author}
  {\bibfnamefont {J.}~\bibnamefont {Simon}},\ }\href@noop {} {\bibfield
  {journal} {\bibinfo  {journal} {arXiv preprint arXiv:1705.07475}\ } (\bibinfo
  {year} {2017})}\BibitemShut {NoStop}%
\bibitem [{\citenamefont {Chen}\ \emph {et~al.}(2016)\citenamefont {Chen},
  \citenamefont {Xiao},\ and\ \citenamefont {Chan}}]{chen2016photonic}%
  \BibitemOpen
  \bibfield  {author} {\bibinfo {author} {\bibfnamefont {W.-J.}\ \bibnamefont
  {Chen}}, \bibinfo {author} {\bibfnamefont {M.}~\bibnamefont {Xiao}}, \ and\
  \bibinfo {author} {\bibfnamefont {C.~T.}\ \bibnamefont {Chan}},\ }\href@noop
  {} {\bibfield  {journal} {\bibinfo  {journal} {Nature communications}\
  }\textbf {\bibinfo {volume} {7}},\ \bibinfo {pages} {13038} (\bibinfo {year}
  {2016})}\BibitemShut {NoStop}%
\bibitem [{\citenamefont {Armitage}\ \emph
  {et~al.}(2018{\natexlab{b}})\citenamefont {Armitage}, \citenamefont {Mele},\
  and\ \citenamefont {Vishwanath}}]{armitage2018}%
  \BibitemOpen
  \bibfield  {author} {\bibinfo {author} {\bibfnamefont {N.~P.}\ \bibnamefont
  {Armitage}}, \bibinfo {author} {\bibfnamefont {E.~J.}\ \bibnamefont {Mele}},
  \ and\ \bibinfo {author} {\bibfnamefont {A.}~\bibnamefont {Vishwanath}},\
  }\href@noop {} {\bibfield  {journal} {\bibinfo  {journal} {Rev. Mod. Phys.}\
  }\textbf {\bibinfo {volume} {90}},\ \bibinfo {pages} {015001} (\bibinfo
  {year} {2018}{\natexlab{b}})}\BibitemShut {NoStop}%
\bibitem [{\citenamefont {Soluyanov}\ \emph {et~al.}(2015)\citenamefont
  {Soluyanov}, \citenamefont {Gresch}, \citenamefont {Wang}, \citenamefont
  {Wu}, \citenamefont {Troyer}, \citenamefont {Dai},\ and\ \citenamefont
  {Bernevig}}]{soluyanov2015type}%
  \BibitemOpen
  \bibfield  {author} {\bibinfo {author} {\bibfnamefont {A.~A.}\ \bibnamefont
  {Soluyanov}}, \bibinfo {author} {\bibfnamefont {D.}~\bibnamefont {Gresch}},
  \bibinfo {author} {\bibfnamefont {Z.}~\bibnamefont {Wang}}, \bibinfo {author}
  {\bibfnamefont {Q.}~\bibnamefont {Wu}}, \bibinfo {author} {\bibfnamefont
  {M.}~\bibnamefont {Troyer}}, \bibinfo {author} {\bibfnamefont
  {X.}~\bibnamefont {Dai}}, \ and\ \bibinfo {author} {\bibfnamefont {B.~A.}\
  \bibnamefont {Bernevig}},\ }\href@noop {} {\bibfield  {journal} {\bibinfo
  {journal} {Nature}\ }\textbf {\bibinfo {volume} {527}},\ \bibinfo {pages}
  {495} (\bibinfo {year} {2015})}\BibitemShut {NoStop}%
\bibitem [{\citenamefont {Yang}\ \emph {et~al.}(2017)\citenamefont {Yang},
  \citenamefont {Guo}, \citenamefont {Tremain}, \citenamefont {Barr},
  \citenamefont {Gao}, \citenamefont {Liu}, \citenamefont {B{\'e}ri},
  \citenamefont {Xiang}, \citenamefont {Fan}, \citenamefont {Hibbins} \emph
  {et~al.}}]{yang2017direct}%
  \BibitemOpen
  \bibfield  {author} {\bibinfo {author} {\bibfnamefont {B.}~\bibnamefont
  {Yang}}, \bibinfo {author} {\bibfnamefont {Q.}~\bibnamefont {Guo}}, \bibinfo
  {author} {\bibfnamefont {B.}~\bibnamefont {Tremain}}, \bibinfo {author}
  {\bibfnamefont {L.~E.}\ \bibnamefont {Barr}}, \bibinfo {author}
  {\bibfnamefont {W.}~\bibnamefont {Gao}}, \bibinfo {author} {\bibfnamefont
  {H.}~\bibnamefont {Liu}}, \bibinfo {author} {\bibfnamefont {B.}~\bibnamefont
  {B{\'e}ri}}, \bibinfo {author} {\bibfnamefont {Y.}~\bibnamefont {Xiang}},
  \bibinfo {author} {\bibfnamefont {D.}~\bibnamefont {Fan}}, \bibinfo {author}
  {\bibfnamefont {A.~P.}\ \bibnamefont {Hibbins}},  \emph {et~al.},\
  }\href@noop {} {\bibfield  {journal} {\bibinfo  {journal} {Nature
  Communications}\ }\textbf {\bibinfo {volume} {8}},\ \bibinfo {pages} {97}
  (\bibinfo {year} {2017})}\BibitemShut {NoStop}%
\bibitem [{\citenamefont {Albert}\ \emph {et~al.}(2015)\citenamefont {Albert},
  \citenamefont {Glazman},\ and\ \citenamefont
  {Jiang}}]{albert2015topological}%
  \BibitemOpen
  \bibfield  {author} {\bibinfo {author} {\bibfnamefont {V.~V.}\ \bibnamefont
  {Albert}}, \bibinfo {author} {\bibfnamefont {L.~I.}\ \bibnamefont {Glazman}},
  \ and\ \bibinfo {author} {\bibfnamefont {L.}~\bibnamefont {Jiang}},\
  }\href@noop {} {\bibfield  {journal} {\bibinfo  {journal} {Physical review
  letters}\ }\textbf {\bibinfo {volume} {114}},\ \bibinfo {pages} {173902}
  (\bibinfo {year} {2015})}\BibitemShut {NoStop}%
\bibitem [{\citenamefont {Lee}\ and\ \citenamefont
  {Thomale}(2017)}]{lee2017topolectrical}%
  \BibitemOpen
  \bibfield  {author} {\bibinfo {author} {\bibfnamefont {C.~H.}\ \bibnamefont
  {Lee}}\ and\ \bibinfo {author} {\bibfnamefont {R.}~\bibnamefont {Thomale}},\
  }\href@noop {} {\bibfield  {journal} {\bibinfo  {journal} {arXiv preprint
  arXiv:1705.01077}\ } (\bibinfo {year} {2017})}\BibitemShut {NoStop}%
\bibitem [{\citenamefont {Luo}\ \emph {et~al.}(2018)\citenamefont {Luo},
  \citenamefont {Yu},\ and\ \citenamefont {Weng}}]{luo2018topological}%
  \BibitemOpen
  \bibfield  {author} {\bibinfo {author} {\bibfnamefont {K.}~\bibnamefont
  {Luo}}, \bibinfo {author} {\bibfnamefont {R.}~\bibnamefont {Yu}}, \ and\
  \bibinfo {author} {\bibfnamefont {H.}~\bibnamefont {Weng}},\ }\href@noop {}
  {\bibfield  {journal} {\bibinfo  {journal} {arXiv preprint arXiv:1801.05581}\
  } (\bibinfo {year} {2018})}\BibitemShut {NoStop}%
\bibitem [{\citenamefont {Yang}\ \emph {et~al.}(2018)\citenamefont {Yang},
  \citenamefont {Gao}, \citenamefont {Xue}, \citenamefont {Zhang},
  \citenamefont {He}, \citenamefont {Yang}, \citenamefont {Singh},
  \citenamefont {Chong}, \citenamefont {Zhang},\ and\ \citenamefont
  {Chen}}]{yang2018realization}%
  \BibitemOpen
  \bibfield  {author} {\bibinfo {author} {\bibfnamefont {Y.}~\bibnamefont
  {Yang}}, \bibinfo {author} {\bibfnamefont {Z.}~\bibnamefont {Gao}}, \bibinfo
  {author} {\bibfnamefont {H.}~\bibnamefont {Xue}}, \bibinfo {author}
  {\bibfnamefont {L.}~\bibnamefont {Zhang}}, \bibinfo {author} {\bibfnamefont
  {M.}~\bibnamefont {He}}, \bibinfo {author} {\bibfnamefont {Z.}~\bibnamefont
  {Yang}}, \bibinfo {author} {\bibfnamefont {R.}~\bibnamefont {Singh}},
  \bibinfo {author} {\bibfnamefont {Y.}~\bibnamefont {Chong}}, \bibinfo
  {author} {\bibfnamefont {B.}~\bibnamefont {Zhang}}, \ and\ \bibinfo {author}
  {\bibfnamefont {H.}~\bibnamefont {Chen}},\ }\href@noop {} {\bibfield
  {journal} {\bibinfo  {journal} {arXiv preprint arXiv:1804.03595}\ } (\bibinfo
  {year} {2018})}\BibitemShut {NoStop}%
\bibitem [{\citenamefont {Nielsen}\ and\ \citenamefont
  {Ninomiya}(1981)}]{nielsen1981no}%
  \BibitemOpen
  \bibfield  {author} {\bibinfo {author} {\bibfnamefont {H.~B.}\ \bibnamefont
  {Nielsen}}\ and\ \bibinfo {author} {\bibfnamefont {M.}~\bibnamefont
  {Ninomiya}},\ }\href@noop {} {\bibfield  {journal} {\bibinfo  {journal}
  {Physics Letters B}\ }\textbf {\bibinfo {volume} {105}},\ \bibinfo {pages}
  {219} (\bibinfo {year} {1981})}\BibitemShut {NoStop}%
\bibitem [{\citenamefont {Hosur}(2012)}]{Hosur2012}%
  \BibitemOpen
  \bibfield  {author} {\bibinfo {author} {\bibfnamefont {P.}~\bibnamefont
  {Hosur}},\ }\href@noop {} {\bibfield  {journal} {\bibinfo  {journal} {Phys.
  Rev. B}\ }\textbf {\bibinfo {volume} {86}},\ \bibinfo {pages} {195102}
  (\bibinfo {year} {2012})}\BibitemShut {NoStop}%
\bibitem [{\citenamefont {Anderson}\ \emph {et~al.}(2016)\citenamefont
  {Anderson}, \citenamefont {Ma}, \citenamefont {Owens}, \citenamefont
  {Schuster},\ and\ \citenamefont {Simon}}]{anderson2016engineering}%
  \BibitemOpen
  \bibfield  {author} {\bibinfo {author} {\bibfnamefont {B.~M.}\ \bibnamefont
  {Anderson}}, \bibinfo {author} {\bibfnamefont {R.}~\bibnamefont {Ma}},
  \bibinfo {author} {\bibfnamefont {C.}~\bibnamefont {Owens}}, \bibinfo
  {author} {\bibfnamefont {D.~I.}\ \bibnamefont {Schuster}}, \ and\ \bibinfo
  {author} {\bibfnamefont {J.}~\bibnamefont {Simon}},\ }\href@noop {}
  {\bibfield  {journal} {\bibinfo  {journal} {Physical Review X}\ }\textbf
  {\bibinfo {volume} {6}},\ \bibinfo {pages} {041043} (\bibinfo {year}
  {2016})}\BibitemShut {NoStop}%
\bibitem [{\citenamefont {Roy}\ \emph {et~al.}(2017)\citenamefont {Roy},
  \citenamefont {Goswami},\ and\ \citenamefont
  {Juri{\v{c}}i{\'c}}}]{roy2017interacting}%
  \BibitemOpen
  \bibfield  {author} {\bibinfo {author} {\bibfnamefont {B.}~\bibnamefont
  {Roy}}, \bibinfo {author} {\bibfnamefont {P.}~\bibnamefont {Goswami}}, \ and\
  \bibinfo {author} {\bibfnamefont {V.}~\bibnamefont {Juri{\v{c}}i{\'c}}},\
  }\href@noop {} {\bibfield  {journal} {\bibinfo  {journal} {Physical Review
  B}\ }\textbf {\bibinfo {volume} {95}},\ \bibinfo {pages} {201102} (\bibinfo
  {year} {2017})}\BibitemShut {NoStop}%
\bibitem [{\citenamefont {Chan}\ and\ \citenamefont {Liu}(2017)}]{Chan2017}%
  \BibitemOpen
  \bibfield  {author} {\bibinfo {author} {\bibfnamefont {C.}~\bibnamefont
  {Chan}}\ and\ \bibinfo {author} {\bibfnamefont {X.-J.}\ \bibnamefont {Liu}},\
  }\href@noop {} {\bibfield  {journal} {\bibinfo  {journal} {Phys. Rev. Lett.}\
  }\textbf {\bibinfo {volume} {118}},\ \bibinfo {pages} {207002} (\bibinfo
  {year} {2017})}\BibitemShut {NoStop}%
\bibitem [{\citenamefont {Buividovich}\ and\ \citenamefont
  {Puhr}(2014)}]{buividovich2014lattice}%
  \BibitemOpen
  \bibfield  {author} {\bibinfo {author} {\bibfnamefont {P.}~\bibnamefont
  {Buividovich}}\ and\ \bibinfo {author} {\bibfnamefont {M.}~\bibnamefont
  {Puhr}},\ }\href@noop {} {\bibfield  {journal} {\bibinfo  {journal} {arXiv
  preprint arXiv:1410.6704}\ } (\bibinfo {year} {2014})}\BibitemShut {NoStop}%
\bibitem [{\citenamefont {{\L}{\k{a}}cki}\ \emph {et~al.}(2016)\citenamefont
  {{\L}{\k{a}}cki}, \citenamefont {Pichler}, \citenamefont {Sterdyniak},
  \citenamefont {Lyras}, \citenamefont {Lembessis}, \citenamefont {Al-Dossary},
  \citenamefont {Budich},\ and\ \citenamefont {Zoller}}]{lkacki2016quantum}%
  \BibitemOpen
  \bibfield  {author} {\bibinfo {author} {\bibfnamefont {M.}~\bibnamefont
  {{\L}{\k{a}}cki}}, \bibinfo {author} {\bibfnamefont {H.}~\bibnamefont
  {Pichler}}, \bibinfo {author} {\bibfnamefont {A.}~\bibnamefont {Sterdyniak}},
  \bibinfo {author} {\bibfnamefont {A.}~\bibnamefont {Lyras}}, \bibinfo
  {author} {\bibfnamefont {V.~E.}\ \bibnamefont {Lembessis}}, \bibinfo {author}
  {\bibfnamefont {O.}~\bibnamefont {Al-Dossary}}, \bibinfo {author}
  {\bibfnamefont {J.~C.}\ \bibnamefont {Budich}}, \ and\ \bibinfo {author}
  {\bibfnamefont {P.}~\bibnamefont {Zoller}},\ }\href@noop {} {\bibfield
  {journal} {\bibinfo  {journal} {Physical Review A}\ }\textbf {\bibinfo
  {volume} {93}},\ \bibinfo {pages} {013604} (\bibinfo {year}
  {2016})}\BibitemShut {NoStop}%
\bibitem [{\citenamefont {Gaunt}\ and\ \citenamefont
  {Hadzibabic}(2012)}]{gaunt2012robust}%
  \BibitemOpen
  \bibfield  {author} {\bibinfo {author} {\bibfnamefont {A.~L.}\ \bibnamefont
  {Gaunt}}\ and\ \bibinfo {author} {\bibfnamefont {Z.}~\bibnamefont
  {Hadzibabic}},\ }\href@noop {} {\bibfield  {journal} {\bibinfo  {journal}
  {Scientific reports}\ }\textbf {\bibinfo {volume} {2}},\ \bibinfo {pages}
  {721} (\bibinfo {year} {2012})}\BibitemShut {NoStop}%
\bibitem [{\citenamefont {Preiss}\ \emph {et~al.}(2015)\citenamefont {Preiss},
  \citenamefont {Ma}, \citenamefont {Tai}, \citenamefont {Lukin}, \citenamefont
  {Rispoli}, \citenamefont {Zupancic}, \citenamefont {Lahini}, \citenamefont
  {Islam},\ and\ \citenamefont {Greiner}}]{preiss2015strongly}%
  \BibitemOpen
  \bibfield  {author} {\bibinfo {author} {\bibfnamefont {P.~M.}\ \bibnamefont
  {Preiss}}, \bibinfo {author} {\bibfnamefont {R.}~\bibnamefont {Ma}}, \bibinfo
  {author} {\bibfnamefont {M.~E.}\ \bibnamefont {Tai}}, \bibinfo {author}
  {\bibfnamefont {A.}~\bibnamefont {Lukin}}, \bibinfo {author} {\bibfnamefont
  {M.}~\bibnamefont {Rispoli}}, \bibinfo {author} {\bibfnamefont
  {P.}~\bibnamefont {Zupancic}}, \bibinfo {author} {\bibfnamefont
  {Y.}~\bibnamefont {Lahini}}, \bibinfo {author} {\bibfnamefont
  {R.}~\bibnamefont {Islam}}, \ and\ \bibinfo {author} {\bibfnamefont
  {M.}~\bibnamefont {Greiner}},\ }\href@noop {} {\bibfield  {journal} {\bibinfo
   {journal} {Science}\ }\textbf {\bibinfo {volume} {347}},\ \bibinfo {pages}
  {1229} (\bibinfo {year} {2015})}\BibitemShut {NoStop}%
\bibitem [{\citenamefont {Mukherjee}\ \emph {et~al.}(2017)\citenamefont
  {Mukherjee}, \citenamefont {Yan}, \citenamefont {Patel}, \citenamefont
  {Hadzibabic}, \citenamefont {Yefsah}, \citenamefont {Struck},\ and\
  \citenamefont {Zwierlein}}]{mukherjee2017homogeneous}%
  \BibitemOpen
  \bibfield  {author} {\bibinfo {author} {\bibfnamefont {B.}~\bibnamefont
  {Mukherjee}}, \bibinfo {author} {\bibfnamefont {Z.}~\bibnamefont {Yan}},
  \bibinfo {author} {\bibfnamefont {P.~B.}\ \bibnamefont {Patel}}, \bibinfo
  {author} {\bibfnamefont {Z.}~\bibnamefont {Hadzibabic}}, \bibinfo {author}
  {\bibfnamefont {T.}~\bibnamefont {Yefsah}}, \bibinfo {author} {\bibfnamefont
  {J.}~\bibnamefont {Struck}}, \ and\ \bibinfo {author} {\bibfnamefont {M.~W.}\
  \bibnamefont {Zwierlein}},\ }\href@noop {} {\bibfield  {journal} {\bibinfo
  {journal} {Physical review letters}\ }\textbf {\bibinfo {volume} {118}},\
  \bibinfo {pages} {123401} (\bibinfo {year} {2017})}\BibitemShut {NoStop}%
\bibitem [{\citenamefont {Stuhl}\ \emph {et~al.}(2015)\citenamefont {Stuhl},
  \citenamefont {Lu}, \citenamefont {Aycock}, \citenamefont {Genkina},\ and\
  \citenamefont {Spielman}}]{stuhl2015visualizing}%
  \BibitemOpen
  \bibfield  {author} {\bibinfo {author} {\bibfnamefont {B.}~\bibnamefont
  {Stuhl}}, \bibinfo {author} {\bibfnamefont {H.-I.}\ \bibnamefont {Lu}},
  \bibinfo {author} {\bibfnamefont {L.}~\bibnamefont {Aycock}}, \bibinfo
  {author} {\bibfnamefont {D.}~\bibnamefont {Genkina}}, \ and\ \bibinfo
  {author} {\bibfnamefont {I.}~\bibnamefont {Spielman}},\ }\href@noop {}
  {\bibfield  {journal} {\bibinfo  {journal} {Science}\ }\textbf {\bibinfo
  {volume} {349}},\ \bibinfo {pages} {1514} (\bibinfo {year}
  {2015})}\BibitemShut {NoStop}%
\bibitem [{\citenamefont {Mancini}\ \emph {et~al.}(2015)\citenamefont
  {Mancini}, \citenamefont {Pagano}, \citenamefont {Cappellini}, \citenamefont
  {Livi}, \citenamefont {Rider}, \citenamefont {Catani}, \citenamefont {Sias},
  \citenamefont {Zoller}, \citenamefont {Inguscio}, \citenamefont {Dalmonte}
  \emph {et~al.}}]{mancini2015observation}%
  \BibitemOpen
  \bibfield  {author} {\bibinfo {author} {\bibfnamefont {M.}~\bibnamefont
  {Mancini}}, \bibinfo {author} {\bibfnamefont {G.}~\bibnamefont {Pagano}},
  \bibinfo {author} {\bibfnamefont {G.}~\bibnamefont {Cappellini}}, \bibinfo
  {author} {\bibfnamefont {L.}~\bibnamefont {Livi}}, \bibinfo {author}
  {\bibfnamefont {M.}~\bibnamefont {Rider}}, \bibinfo {author} {\bibfnamefont
  {J.}~\bibnamefont {Catani}}, \bibinfo {author} {\bibfnamefont
  {C.}~\bibnamefont {Sias}}, \bibinfo {author} {\bibfnamefont {P.}~\bibnamefont
  {Zoller}}, \bibinfo {author} {\bibfnamefont {M.}~\bibnamefont {Inguscio}},
  \bibinfo {author} {\bibfnamefont {M.}~\bibnamefont {Dalmonte}},  \emph
  {et~al.},\ }\href@noop {} {\bibfield  {journal} {\bibinfo  {journal}
  {Science}\ }\textbf {\bibinfo {volume} {349}},\ \bibinfo {pages} {1510}
  (\bibinfo {year} {2015})}\BibitemShut {NoStop}%
\bibitem [{\citenamefont {Celi}\ \emph {et~al.}(2014)\citenamefont {Celi},
  \citenamefont {Massignan}, \citenamefont {Ruseckas}, \citenamefont {Goldman},
  \citenamefont {Spielman}, \citenamefont {Juzeli{\=u}nas},\ and\ \citenamefont
  {Lewenstein}}]{celi2014synthetic}%
  \BibitemOpen
  \bibfield  {author} {\bibinfo {author} {\bibfnamefont {A.}~\bibnamefont
  {Celi}}, \bibinfo {author} {\bibfnamefont {P.}~\bibnamefont {Massignan}},
  \bibinfo {author} {\bibfnamefont {J.}~\bibnamefont {Ruseckas}}, \bibinfo
  {author} {\bibfnamefont {N.}~\bibnamefont {Goldman}}, \bibinfo {author}
  {\bibfnamefont {I.~B.}\ \bibnamefont {Spielman}}, \bibinfo {author}
  {\bibfnamefont {G.}~\bibnamefont {Juzeli{\=u}nas}}, \ and\ \bibinfo {author}
  {\bibfnamefont {M.}~\bibnamefont {Lewenstein}},\ }\href@noop {} {\bibfield
  {journal} {\bibinfo  {journal} {Physical review letters}\ }\textbf {\bibinfo
  {volume} {112}},\ \bibinfo {pages} {043001} (\bibinfo {year}
  {2014})}\BibitemShut {NoStop}%
\bibitem [{\citenamefont {Biswas}\ and\ \citenamefont
  {Son}(2016)}]{biswas2016fractional}%
  \BibitemOpen
  \bibfield  {author} {\bibinfo {author} {\bibfnamefont {R.~R.}\ \bibnamefont
  {Biswas}}\ and\ \bibinfo {author} {\bibfnamefont {D.~T.}\ \bibnamefont
  {Son}},\ }\href@noop {} {\bibfield  {journal} {\bibinfo  {journal}
  {Proceedings of the National Academy of Sciences}\ }\textbf {\bibinfo
  {volume} {113}},\ \bibinfo {pages} {8636} (\bibinfo {year}
  {2016})}\BibitemShut {NoStop}%
\bibitem [{\citenamefont {He}\ \emph {et~al.}(2012)\citenamefont {He},
  \citenamefont {Moore},\ and\ \citenamefont {Varma}}]{He2012}%
  \BibitemOpen
  \bibfield  {author} {\bibinfo {author} {\bibfnamefont {Y.}~\bibnamefont
  {He}}, \bibinfo {author} {\bibfnamefont {J.}~\bibnamefont {Moore}}, \ and\
  \bibinfo {author} {\bibfnamefont {C.~M.}\ \bibnamefont {Varma}},\ }\href@noop
  {} {\bibfield  {journal} {\bibinfo  {journal} {Phys. Rev. B}\ }\textbf
  {\bibinfo {volume} {85}},\ \bibinfo {pages} {155106} (\bibinfo {year}
  {2012})}\BibitemShut {NoStop}%
\bibitem [{\citenamefont {Bernevig}(2015)}]{Bernevig2015}%
  \BibitemOpen
  \bibfield  {author} {\bibinfo {author} {\bibfnamefont {B.~A.}\ \bibnamefont
  {Bernevig}},\ }\href@noop {} {\bibfield  {journal} {\bibinfo  {journal}
  {Nature Physics}\ }\textbf {\bibinfo {volume} {11}},\ \bibinfo {pages} {698
  EP } (\bibinfo {year} {2015})}\BibitemShut {NoStop}%
\end{thebibliography}%
\onecolumngrid
\renewcommand\appendixname{Supplement}
\appendix
\clearpage
{\centering\Huge Supplementary Information}
\section{WEYL HAMILTONIAN}
\label{SI:WeylHam}
Because the synthetic flux penetrating each plaquette in any principal plane ($x-y$, $x-z$, or $y-z$) is $\pi$, we can write a tight-binding Hamiltonian: 
$$
\begin{aligned}
H=\sum_{i,j}H_{ij}c^\dagger_i c_j =\sum_i\varepsilon_0\left(n_{iA}+n_{iB}\right)\\
-\sum_{i}t_0\left(c^\dagger_A(\bm{r}_i) c_B(\bm{r}_i+\bm{\hat{x}}) +c^\dagger_A(\bm{r}_i) c_B(\bm{r}_i-\bm{\hat{x}})+H.C.\right)\\
-\sum_{i}t_0\left(c^\dagger_A(\bm{r}_i) c_B(\bm{r}_i+\bm{\hat{y}}) 
-c^\dagger_A(\bm{r}_i) c_B(\bm{r}_i-\bm{\hat{y}})+H.C.\right)\\
-\sum_{i}t_0\left(c^\dagger_A(\bm{r}_i) c_A(\bm{r}_i+\bm{\hat{z}})-c^\dagger_B(\bm{r}_i) c_B(\bm{r}_i+\bm{\hat{z}})+H.C.\right)
\end{aligned}
$$

In the basis $\left(c_{\bm{k}A},c_{\bm{k}B}\right)$ the Bloch Hamiltonian reads: 
$$\mathcal{H}(\bm{k})=\varepsilon_0  +\bm{h}(\bm{k})\cdot \bm{\sigma}$$
Where $\bm{h}(\bm{k})=2t_0 \left(-\cos(k_x a),\, \sin(k_y a),\, -\cos(k_z a) \right)$.

The eigenstate energy is therefore:
$$E_\pm(\bm{k})=\varepsilon_0  \pm 2t_0 \sqrt{\cos^2(k_z a)+ \cos^2(k_x a)+\sin^2(k_y a)}$$

It is apparent that the Hamiltonian has 4 Weyl points in the first BZ, all located in the $k_y=0$ plane. The corresponding (pseudo-) spin texture of the two bands is: 
$$\bm{a}_{\pm}(\bm{k})=\frac{\pm\bm{h}(\bm{k})}{\lvert\bm{h}(\bm{k})\rvert}=\left(\mp\cos(k_x a),\, \pm\sin(k_y a),\, \mp\cos(k_z a) \right)$$

For convenience, we let $a=1$ in what follows. The 4 Weyl points are then located at: $(\pm\pi/2,0,\pm\pi/2)$. We further define $k_u=\frac{1}{\sqrt{2}}(k_x+k_y), k_v=\frac{1}{\sqrt{2}}(k_x-k_y)$.

\section{TOPOLOGICAL CIRCUIT THEORY}
\label{SI:Topological Circuit Theory}

In SI~\ref{SI:WeylHam} we derived the Hamiltonian and eigen-spectrum of a tight-binding model with the same connectivity as our circuit. Here we establish a one-on-one correspondence between the tight-binding model and the RF circuit that we actually built and studied. Intuitively, each on-site LC resonator has a oscillation frequency that resembles the on-site energy in a tight-binding model, and the coupling strength (capacitance of the inter-site coupling capacitors) in the circuit controls the tunneling rate. We will see that one can conveniently throw away the capacitor in the on-site LC resonator pair, because of what we call a ``shifting terms'' in an admittance matrix. The rigorous derivation follows:

The mathematics behind a tight-binding Hamiltonian assumes that the Wannier functions are sufficiently localized that the hopping terms between different orbitals (i.e. s- and p- orbitals) can be neglected, while hopping terms between the same orbitals on different sites can be expressed in terms of the overlap integral of the Hamiltonian acting on the two Wannier functions. In the spirit of the tight-binding approximation, we are able to expand the wave-function in the Wannier basis, thereby projecting an uncountably infinite dimensional Hilbert space onto a countably infinite dimensional space, and thus reducing the (partial differential) Schrödinger equation into a matrix eigenvalue equation:
$$\sum_n\mel{m}{H}{n}\bra{n}\ket{\Psi}=E\bra{m}\ket{\Psi}$$
where the state of the system can be expressed by the column vector $\bra{n}\ket{\Psi}$, and the matrix elements $H_{mn}=\mel{m}{H}{n}$. The eigenvalues of the Hamiltonian may be obtained through a secular equation. This draws a mathematical resemblance to a set of coupled oscillators, where the eigen-frequencies are the solutions to the secular equation of the matrix equation of the oscillators. 

We derive the corresponding coupled-LC oscillator circuit through the following steps (Fig.~\ref{fig:setup}): (1) In an $N$-site system, for every site $i$ ($1\leq i\leq N$) we place an inductor of identical inductance $L_{0}$ on each site and connect it to a capacitor $C_{i}$, and with positive terminal labeled $i_+$, negative terminal $i_-$; (2) For inter-site coupling connections between site $i$ and site $j$ ($i\neq j$), we either connect $i_+$ to $j_+$, $i_-$ to $j_-$ with two capacitors of equal value $C_{ij}^+$, or connect $i_+$ to $j_-$ and $i_-$ to $j_+$ (which we call ``braiding'') with two capacitors of equal value $C_{ij}^-$. We then define the value and braiding relation of inter-site capacitors to be
$$C_{ji}=C_{ij}=
    \begin{cases}
      C_{ij}^+, & \text{if braiding is false}\\
      C_{ij}^-, & \text{if braiding is true}
    \end{cases}$$
    
$$\phi_{ji}=-\phi_{ij}=
    \begin{cases}
      0, & \text{if braiding is false}\\
      \pi, & \text{if braiding is true}
    \end{cases}$$

Let the electric potential at $i+$ and $i-$ be $V_{i+}$ and $V_{i-}$, respectively. We can recombine the variables into $U_{i}=(V_{i+}-V_{i-})/2$ and $\bar{V}_{i}=(V_{i+}+V_{i-})/2$. It is straightforward to show that $U_{i}$ and $\bar{V}_{i}$ evolve independently of eachother, with $\bar{V}_{i}$ being a constant voltage that is irrelevant to us (it produces no current in the inductor, and so does not couple to our drive- or pickup- coils). To acquire the equation of motion for $U_{i}$, we let $\bar{V}_{i}=0$ for all $i$, which is equivalent to adding a ground in the ``middle'' of each site, as in the following circuit in Fig.~\ref{fig:circuit_braiding}:

\begin{figure*}
\centering
\includegraphics[width=0.9\textwidth]{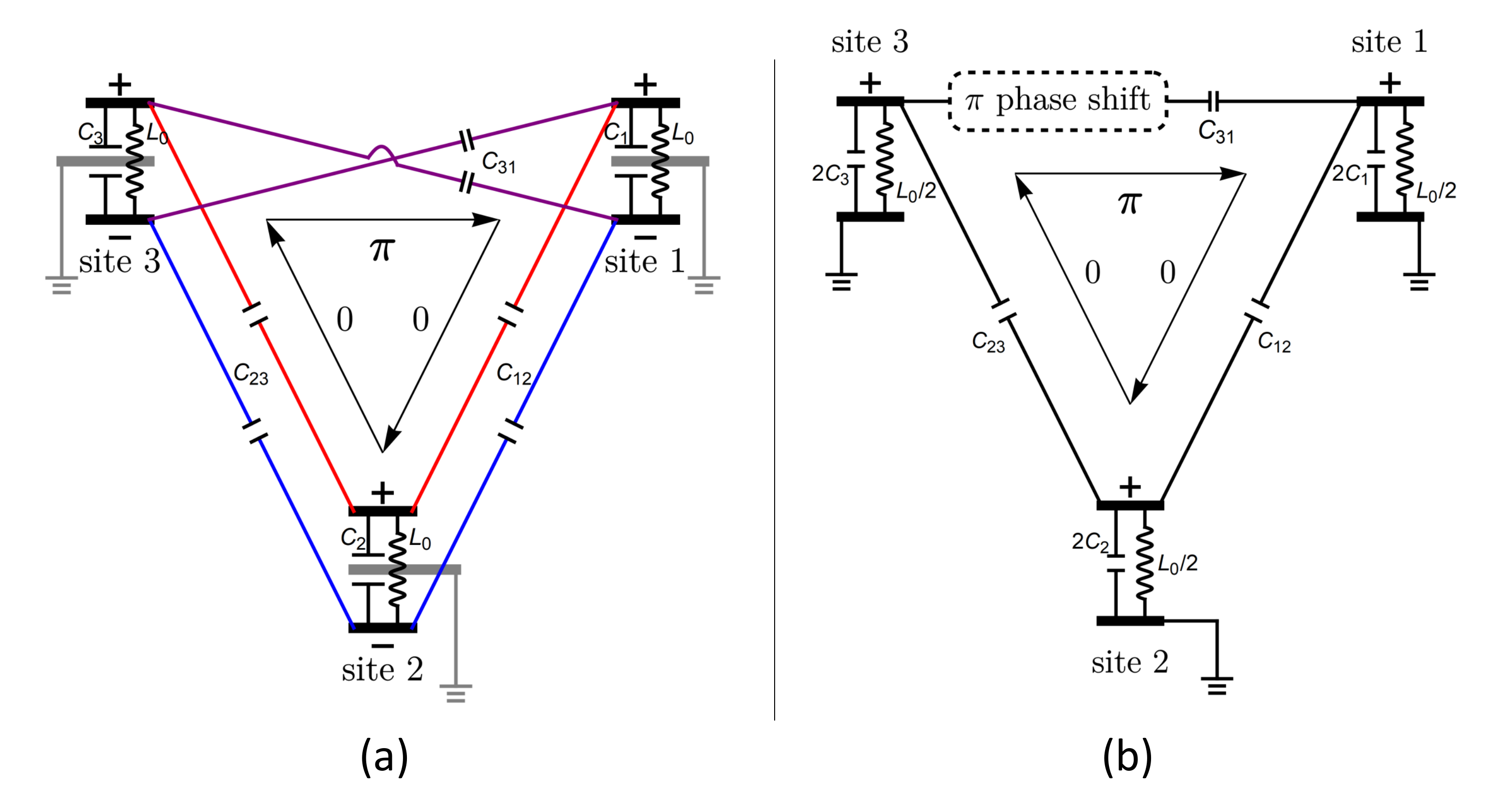}
\caption{\label{fig:circuit_braiding} An example of a 3-site-circuit. In \textbf{(a)} we depict an ideal 3-site system with 2 unbraided connections and 1 braided connection. On every site $i$ ($1\leq i\leq 3$) there is an inductor of identical inductance $L_{0}$ parallel to a capacitor $C_{i}$. For inter-site coupling connections between site $1$ and site $2$/ site $2$ and site $3$, the same-sign poles are connected with two capacitors of equal value $C_{12}$/$C_{23}$, as shown by the red and blue lines (braiding-false), where as between site $3$ and site $1$, the opposite-sign poles are connected with two capacitors of equal value $C_{31}$ shown in purple lines (braiding-true). Adding a virtual ground to the middle of each of the sites as shown in gray, \textbf{(a)} is equivalent to \textbf{(b)}, where we have 3 grounded resonators of component values as shown; The braiding is depicted in the grounded picture with dashed box to indicate the $\pi$ coupling phase.}
\end{figure*}

This is an N-bus circuit which obeys the Kirchoff equations: 
$$I_{i}=\sum_{j=1}^N Y_{ij} U_j$$

Where $Y_{ij}$ is the admittance matrix, whose matrix elements are:
$$Y_{ij}=\begin{cases}
      \frac{2}{i\omega L_0}+2i\omega C_{i} +i\omega \sum_{k=1}^N C_{ik}, & i=j\\
      -i\omega C_{ij} e^{i\phi_{ij}} , & i\neq j
    \end{cases}$$

and $I_{i}$ is the external current injected into site $i$. To compute the eigen-frequencies, we consider an isolated system with zero input current. We then have:
$$\sum_{j=1}^N Y_{ij} U_j=0$$

A non-trivial solution of this equation requires that $det\left(Y\right)=0$, which yields an eigenvalue equation for $\frac{1}{\omega^2}$, whose eigenvectors $U$ correspond to the wave-functions of the tight-binding problem:

$$\frac{1}{\omega^2 } U_i =\sum_{j=1}^N T_{ij}U_j$$

Where the ``resonance'' matrix is:
\begingroup
\renewcommand*{\arraystretch}{1.5}
$$T=L_0
\begin{bmatrix}
    C_{1} + \frac{1}{2}\sum_{k=1}^N C_{1k} & \dots  & -\frac{1}{2} C_{1N}e^{i\phi_{1N}}  \\
     -\frac{1}{2} C_{21}e^{i\phi_{21}}  & \dots  &  -\frac{1}{2} C_{2N}e^{i\phi_{2N}}  \\
    \vdots & \ddots & \vdots \\
    -\frac{1}{2} C_{1N}e^{i\phi_{1N}}  & \dots  & C_{N} + \frac{1}{2}\sum_{k=1}^N C_{Nk}
\end{bmatrix}
$$
\endgroup

For the example circuit shown in Fig.~\ref{fig:circuit_braiding} the ``resonance'' matrix of the 3-site lattice is:

\begingroup
\renewcommand*{\arraystretch}{1.5}
$$T_{e.g.}=L_0
\begin{bmatrix}
    C_{1} + \frac{C_{12}+C{31}}{2} & -\frac{C_{12} }{2}  & +\frac{C_{31}}{2}   \\
     -\frac{ C_{12}}{2}  & C_{2} + \frac{C_{12}+C{23}}{2}  &   -\frac{C_{23}}{2}  \\
    +\frac{C_{31}}{2}   &  -\frac{C_{23}}{2}   & C_{3} + \frac{C_{23}+C{31}}{2}
\end{bmatrix}
$$
\endgroup

We compare the ``resonance'' matrix $T$ to the Hamiltonian on a tight-binding lattice. The resemblance is apparent: The inter-site tunneling strength (off-diagonal terms) pf the latter is proportional to the inter-site connection capacitance in the former, while the braiding of the connection yields either a $0$ phase tunneling (no braiding) or $\pi$ phase tunneling (braiding). The on-site energy of the tight-binding model, however, is proportional to the sum of two terms in the circuit model: (1) The on-site capacitance itself and (2) a sum of capacitances of all coupling capacitors that are connected to this site. The second term is crucial for our implementation and we will refer to it as the ``shifting term.'' The existence of this shifting term is crucial to our observation of Weyl bands because it pushes all eigenvalues of the system away from zero once the self-capacitance is removed, as it is in our experiments.

The most direct mathematical correspondence between the eigen-energies in the tight-binding model and the eigen-frequencies in the circuit model is therefore: 
$$E \propto \frac{1}{\omega^2}$$
Thus, in our experiment we have ``inverted'' the band structure-- this does not impact the topological properties of the Weyl points, nor the linear dispersion in their vicinity.

Note that we can add a global energy offset in a \emph{tight-binding model} without any physical effect, but for the circuit such an offset will distort the band-structure. Nonetheless, the shifting terms are always positive so $T$ remains positive definite. Thus, a small offset will not change the topology of the bands. 

In theory we could invert the bands (send $\omega\rightarrow 1/\omega$) by swapping the roles of inductors and capacitors, thereby realizing a situation where $$E \propto \omega^2$$. This is analogous to switching from a left-handed transmission line to a conventional right-handed transmission line by swapping inductors and capacitors (see SI of~\cite{ningyuan2015time}). We will not do so for technical reasons: 
\begin{enumerate}
\item The photon lifetime in the circuit is limited by inductor resistance, so fewer inductors is favorable.
\item The inductors are physically larger than the capacitors, so employing a configuration with the fewest inductors possible minimizes the physical size of the meta-material.
\item Our measurement procedure relies upon exciting/measuring the local magnetic field of the inductors, so it is simpler if the inductors represent the sites rather than the tunnel-couplers.
\end{enumerate}

\section{Weyl Circuit Model}
\label{SI:WeylCircuitComponents}
Using the circuit mapping introduced in SI~\ref{SI:Topological Circuit Theory}, we assemble an $8\times 8\times 8\time 2$ lattice by stacking printed circuit boards that have connections as shown in Fig.~\ref{fig:setup}b. 

Because no on-site capacitor $C_{i}$ is included, the diagonal terms of the admittance matrix arise solely from the 12 tunneling  capacitors and resulting the ``shifting terms'' described previously. In turn we can write the ``resonance'' matrix as: 

$$T=\frac{L_0}{2}
\begin{bmatrix}
     6 C_0 & \dots  & - C_{1N}e^{i\phi_{1N}}  \\
     - C_{21}e^{i\phi_{21}}  & \dots  &  - C_{2N}e^{i\phi_{2N}}  \\
    \vdots & \ddots & \vdots \\
    - C_{1N}e^{i\phi_{1N}}  & \dots  &  6 C_0
\end{bmatrix}$$

Where the off-diagonal terms only take non-zero values $\mp C_0$ for nearest neighbor connections, and the sign depends on the braiding of the connection. This is in the same form as the tight-binding Hamiltonian in SI~\ref{SI:WeylHam}.

Consequently, in momentum space the equations are:
$$T(\bm{k})= L_0 C_0 \left[ 3 - \cos(k_z a)\sigma_z - \cos(k_x a)\sigma_x + \sin(k_y a)\sigma_y \right]$$

Where $L_0=1 \, \mathrm{mH}$ is the inductance for inductors, and $C_0=1 \times 10^{2}\, \mathrm{pF}$ is the capacitance for capacitors, and we let $a=1$ be the nearest neighbor distance. Theory predicts the frequency of the four Weyl points at $f_{\rm Weyl}=\left(2\pi \sqrt{3L_0 C_0}\right)^{-1}=290  \, \mathrm{kHz}$; the frequency of the bands lies within the frequency range $[f_{min},f_{max}]$, where $f_{max}=\left(2\pi \sqrt{(3-\sqrt{3})L_0 C_0}\right)^{-1}=447 \, \mathrm{kHz} , f_{min}=\left(2\pi \sqrt{(3+\sqrt{3})L_0 C_0}\right)^{-1}=231 \, \mathrm{kHz}$; at the Weyl points $v_{group}= 2\pi \lvert \nabla_{\bm{k}}f_+ \rvert_{\bm{k}\rightarrow Weyl\, point}=\frac{1}{6\sqrt{3 L_0 C_0}}=304\, \mathrm{ms^{-1}}$, all of which we experimentally validate in the body and supplements.

In SI~\ref{SI:Topological Circuit Theory} we established that the mapping from a generalized circuit model to its corresponding tight-binding problem takes $\frac{1}{\omega^2} \rightarrow E$. This swaps the position of the two bands and distorts the shape of the bands due to the non-linearity of the $\omega$ - $E$ mapping. However, this non-linear mapping does not impact the topology underlying the physics. To provide a more intuitive picture of the resemblance between the problems, we linearly expand the frequency of the two bands around the Weyl frequency, to give a approximate but direct mapping from the circuit model to its corresponding tight-binding problem that takes $\omega \rightarrow E$. 

\begin{figure}
\centering
\includegraphics[width=0.49\textwidth]{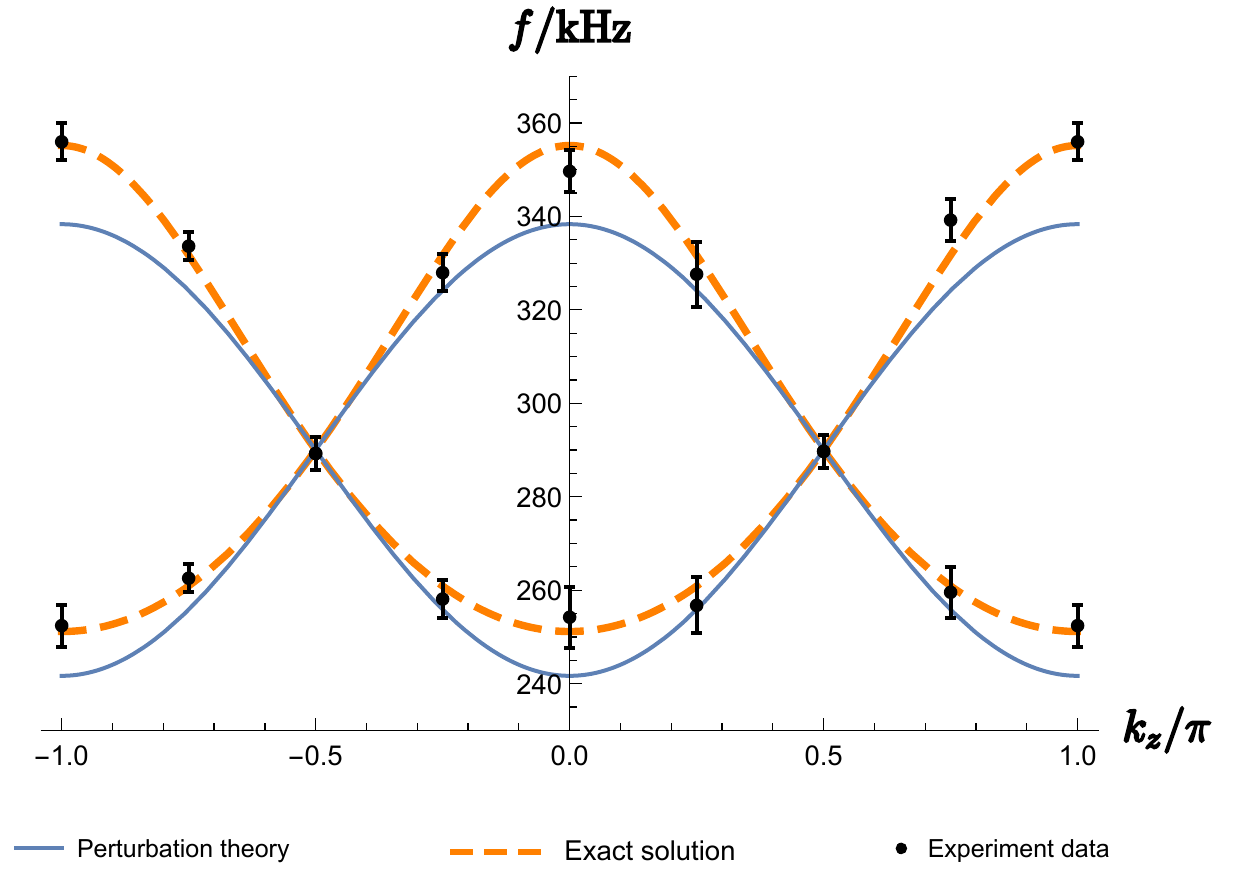}
\caption{\label{fig:theory_vs_experiment} A comparison of the dispersion relation along the $k_x=\pi/2$, $k_y=0$ line in k-space, between the perturbative result of linear expansion around the Weyl frequency (blue solid), the exact solution (orange dashed), and the experimental results (points). The perturbative solution is identical to the tight-binding Hamiltonian and thus exhibits a band-inversion symmetry about the Weyl frequency. The exact solution is slightly distorted\jcmt{, however still topologically equivalent to the tight-binding model-- EQUIVALENCE IS NOT APPARENT FROM THE DATA}. The experimental data are shown in with error bars marking the linewidth of the spectral feature, as observed in Fig~\ref{fig:NWA}d.}
\end{figure} 

This approximation comes about naturally if we consider the nearest-neighbor coupling terms as a perturbation to the on-site resonators. In k-space, the $2\times 2$ resonance matrix may be written: 
$$T(\bm{k})=T_0+\delta T(\bm{k})$$
Where $T_0=\frac{1}{\omega_0^2} \sigma_0$ and $\delta T(\bm{k})=\frac{1}{3\omega_0^2} [- \cos(k_z a)\sigma_z - \cos(k_x a)\sigma_x + \sin(k_y a)\sigma_y ]$, where $\omega_0=1/\sqrt{3 L_0 C_0}=2\pi f_{\rm Weyl}$ is the Weyl angular frequency. 

The eigenvalue equation of $\frac{1}{\omega^2}$ is:
$$\left[\frac{1}{\omega_0^2}+\delta T(\bm{k})\right]\chi(\bm{k})=\frac{1}{\omega^2}\chi(\bm{k})$$ 
$\chi(\bm{k})$ is the spin wavefunction on a certain point in k-space. 

Taking the first-order expansion of the perturbation, we get:
$$\left[\omega_0-\frac{\omega_0^3}{2} \delta T(\bm{k})+\cdots\right]\chi(\bm{k})=\omega\chi(\bm{k})$$

Therefore, taking $\omega$ to be the equivalence of energy in the tight-binding problem, we get the equivalent Hamiltonian under a linear expansion about the Weyl frequency: 
$$\mathcal{H}(\bm{k})=\varepsilon_0  +\bm{h}(\bm{k})\cdot \bm{\sigma}$$
Where $\varepsilon_0=\omega_0=1/\sqrt{3L_0 C_0}$, $\bm{h}(\bm{k})=\frac{\omega_0}{6} \big(\cos(k_x a),$ $\, -\sin(k_y a),\, \cos(k_z a) \big)$. We thereby arrive at the spin-texture and angular frequency of both bands by solving the eigenvectors and eigenvalues of this Hamiltonian. The spin-texture of the upper/lower bands reads: 
$$\bm{a}(\bm{k}) =\pm \frac{\bm{h}(\bm{k})}{|\bm{h}(\bm{k})|} =\pm \frac{\big(\cos(k_x a),\, -\sin(k_y a),\, \cos(k_z a) \big)}{\sqrt{\cos^2(k_x a)+\sin^2(k_y a)+\cos^2(k_z a)}}$$
The corresponding eigen-energies are:
\begin{align*}
\omega(\bm{k}) &=\varepsilon_0 \pm |\bm{h}(\bm{k})|\\
&= \omega_0  \pm \frac{\omega_0}{6} \sqrt{\cos^2(k_x a)+\sin^2(k_y a)+\cos^2(k_z a)}
\end{align*}

\begin{figure*}
\centering
\includegraphics[width=\textwidth]{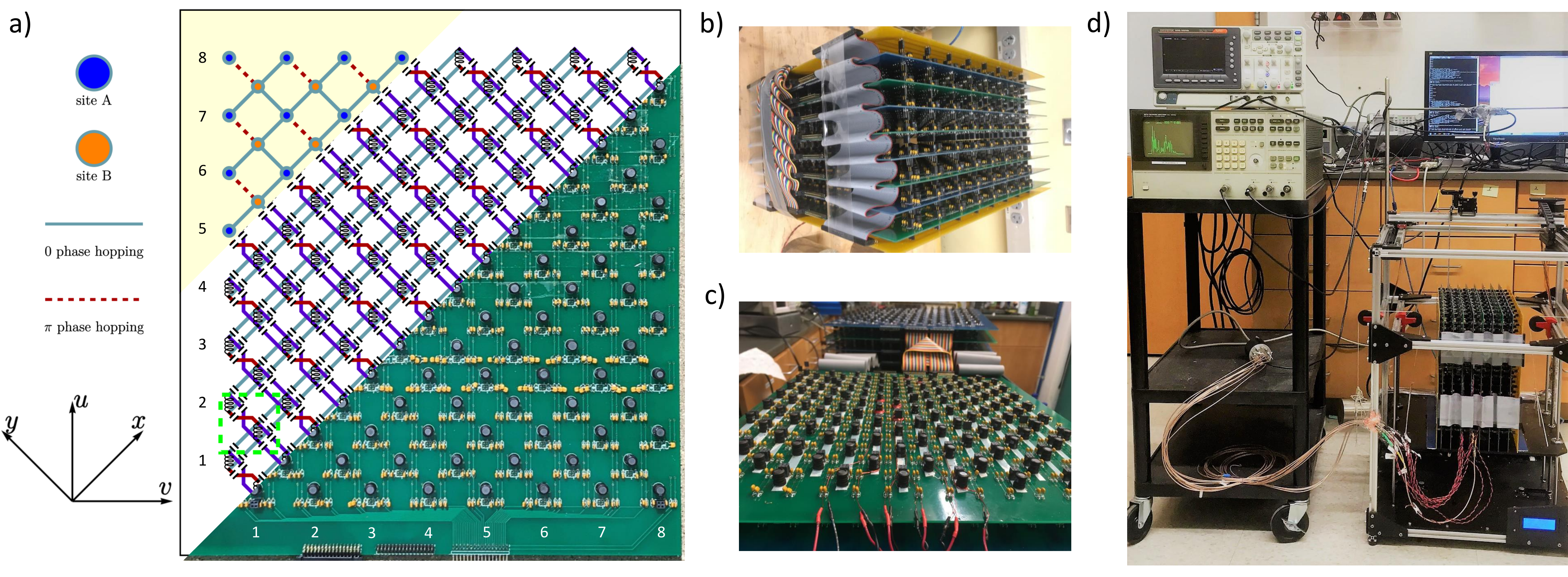}
\caption{\label{fig:whole_board} As shown in \textbf{(a)}, \textbf{(b)} and \textbf{(c)}, the circuit(\ref{SI:Board_design}) is comprised of 8 stacked printed circuit boards (PCBs), each containing two interleaved $8\times 8$ sub-lattices of inductors ($1.02(1)$ mH), capacitively coupled on both ends (by $100(2)$ pF capacitors) to their nearest neighbors. \textbf{(a)} shows (from upper-left to lower-right) the tight-binding-model, the circuit schematic, and an actual photograph of a single circuit board; a single unit cell (outlined by the green dashed square) contains two inductors and 12 capacitors. \textbf{(b)} and \textbf{(c)} shows the stacking of 8 boards; neighboring boards are capacitively coupled together through additional inter-board headers. Periodic boundary conditions(\ref{SI:RealizeBoundary}) are imposed by connecting opposite faces of the full lattice together using ribbon cables, as  shown in \textbf{(b)} and \textbf{(c)}. Input coils are installed around selected inductors as shown in \textbf{(c)}. The entire measurement system is shown in \textbf{(d)}, including a 3D translational apparatus with the circuit board array mounted to its stage and the probe coil mounted to its its arm (in the lower-right corner); an oscilloscope, network analyzer and Radiall 10 way switch on the left (from top to bottom); and a PC, a Raspberry Pi, an Arduino, a TTL switch and a function wave generator in the back. 
See time-lapse video of the data-collecting process at this link: https://youtu.be/LSTsVCF1pfk}
\end{figure*}

In  Fig~\ref{fig:theory_vs_experiment} we compare this result to the exact solution of the ``resonance matrix'' $T$ in SI~\ref{SI:Topological Circuit Theory}, which reads:
$$\omega(\bm{k}) = \omega_0 \left(1 \mp \frac{1}{3} \sqrt{\cos^2(k_x a)+\sin^2(k_y a)+\cos^2(k_z a)} \right)^{-\frac{1}{2}}$$

We note that compared to the tight-binding model in in SI~\ref{SI:WeylHam}, the spin texture $\bm{h}(\bm{k})$ has changed sign due to the sign change of first order linear expansion term, equivalent to the band-swap. Subsequently, the chirality of each of the Weyl points also changes sign. 

\section{EXPERIMENTAL DEVICE AND SETUP}
\label{SI:Device}

The entire experimental setup consists of two essential components: the meta-material 3D LC resonator array itself, and the probing system designed to measure the electrical properties of the meta-material.

\subsection{Circuit Board design}
\label{SI:Board_design}

In practice, we stack 8 identical 2D printed circuit boards (PCBs) in the $\hat{z}$ direction to realize a 3D circuit. On each of the circuit boards are 8-by-8 unit cells with two inductors in each unit cell, as shown in Fig.~\ref{fig:whole_board}a. Within the boards, the nearest neighbor capacitor connections give nearest neighbor tunneling, with sign determined by the braiding of the coupling. The x- and y- connections are not parallel to the PCB edges, but form  $45^{\circ}$ degree angle sto the edges of the PCB. Indeed, the board edges are aligned to the $\hat{u}=\frac{1}{\sqrt{2}}(\hat{x}+\hat{y})$ and $\hat{v}=\frac{1}{\sqrt{2}}(\hat{x}-\hat{y})$ directions. The inductors in adjacent PCBS in the stack are coupled via a pair of capacitors installed on one of the boards, which are then connected to the other board through a header. These header not only establish stable electric connection between the boards, but also provide the mechanical stability that holds the whole 3D circuit without bending under its own weight, whilst remaining hollow enough to permit probe insertion into the bulk. Fig.~\ref{fig:whole_board}b/c shows the assembled board stack/stack opened up between two adjacent boards.

In addition, we design the boundary conditions as follows:
\begin{enumerate}
\item Periodic boundary conditions are applied along the $\hat{v}$ direction with traces on the board;
\item The choice of periodic- or open- boundaries is implemented using external wire connections.
\item Periodic boundary conditions are applied on the $\hat{z}$ direction by adding two connector boards on the top and bottom of the stack of boards, with a cable band connecting the two boards (For details about connection see SI~\ref{SI:RealizeBoundary}). 
\end{enumerate}

We design the edge of the circuit boards along the $v$ and $u$ directions instead of $x$ and $y$ directions to enable us to see topologically protected surface modes in the slab geometry. Since the Fermi Arcs connect projections of Weyl points onto the surface Brillouin zone with non-zero net chirality, there will not be any topologically protected surface states on $\hat{x}-\hat{y}$, $\hat{x}-\hat{z}$ or $\hat{y}-\hat{z}$ oriented surfaces (See SI~\ref{SI:Surface_orientation}).  

Other important design considerations implemented in our experimental setup include:
\begin{enumerate}
\item The absence of on-site capacitors, thus making $C_{onsite}$ zero (see~\cite{ningyuan2015time} for details); this simplifies the design and reduces the component-disorder-sensitivity of the system by maximizing the energy-scale of dynamics. This simplification does not impact the structure of the Weyl points due to the ``shifting terms'' in the admittance matrix (See SI~\ref{SI:Topological Circuit Theory}).
\item We employ low -ariance components to minimize the disorder; this disorder does not impact the existence of the (topologically immune) Weyl points~\cite{armitage2018}, but breaks the translational invariance that we assume in our band-structure reconstruction and theoretical calculations.
\item We employ low resistance/high Q inductors so that excitations can explore the full system before they are damped out, ensuring that our band-structure resolution is limited not by loss, but by finite-system size.
\item We employ braided inter-site connections to realize $\pi$-phase excitation hopping.
\item We include external connectors enabling us to choose between periodic- and open- boundary conditions.
\end{enumerate}

\begin{figure}
\centering
\includegraphics[width=0.42\textwidth]{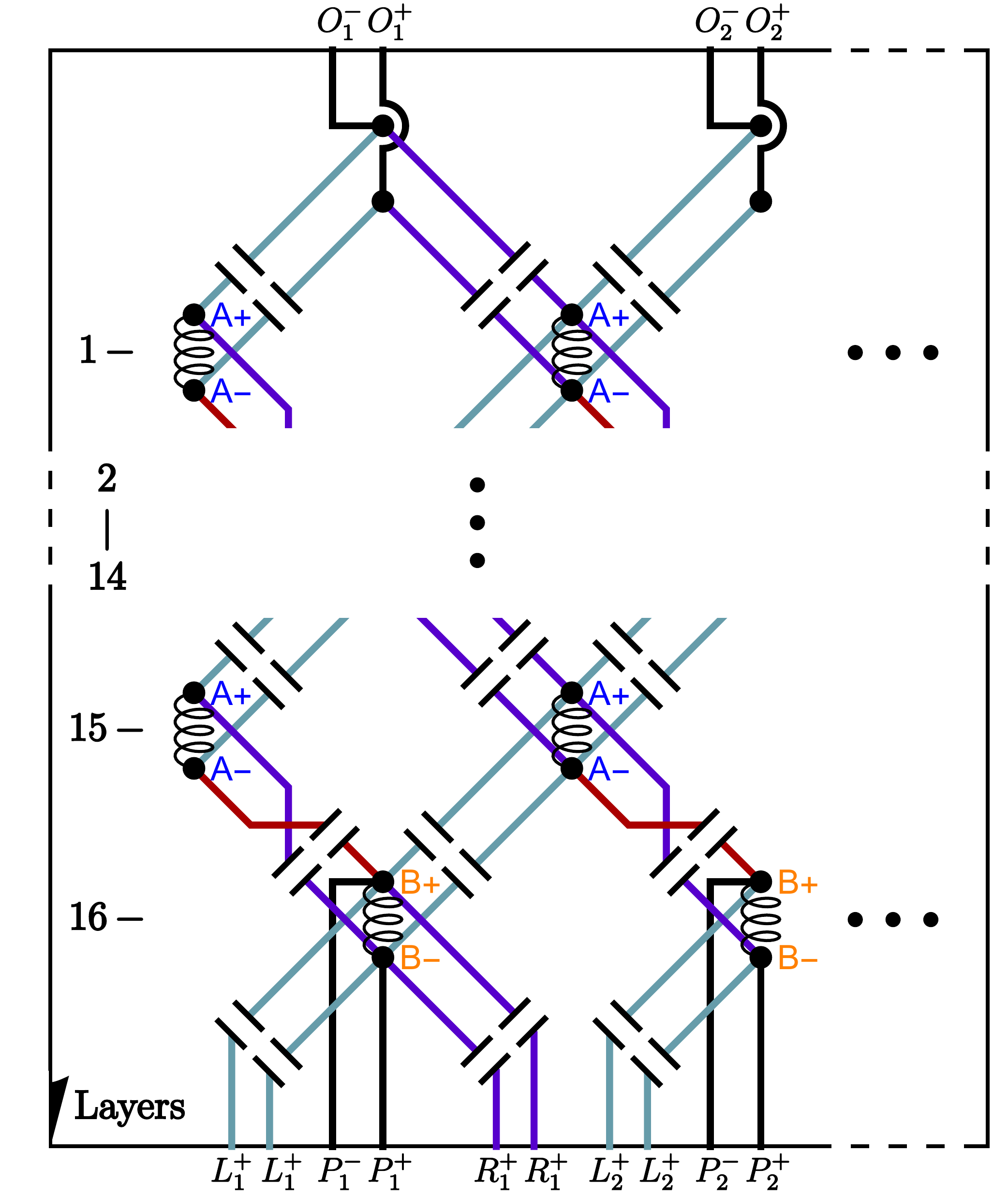}
\caption{\label{fig:boundary_v} Single-board schematic for the optional boundary condition connections in the $v$ dimension, which is the vertical direction as shown in figure with the layers enumerated in the side-bar. To realize periodic boundary conditions, we connect the corresponding $O_i^{\pm}$ to $P_i^{\pm}$ with a cable; to realized open boundary condition with 16 layers, we short-circuit each pair of $O_i^{+}$ to $O_i^{-}$,  $L_i^{+}$ to  $L_i^{-}$ and $R_i^{+}$ to  $R_i^{-}$; to realized open boundary condition with 15 layers, we short circuit each pair of  $O_i^{+}$ to  $O_i^{-}$ and $P_i^{+}$ to  $P_i^{-}$.}
\end{figure}

\begin{figure}
\centering
\includegraphics[width=0.42\textwidth]{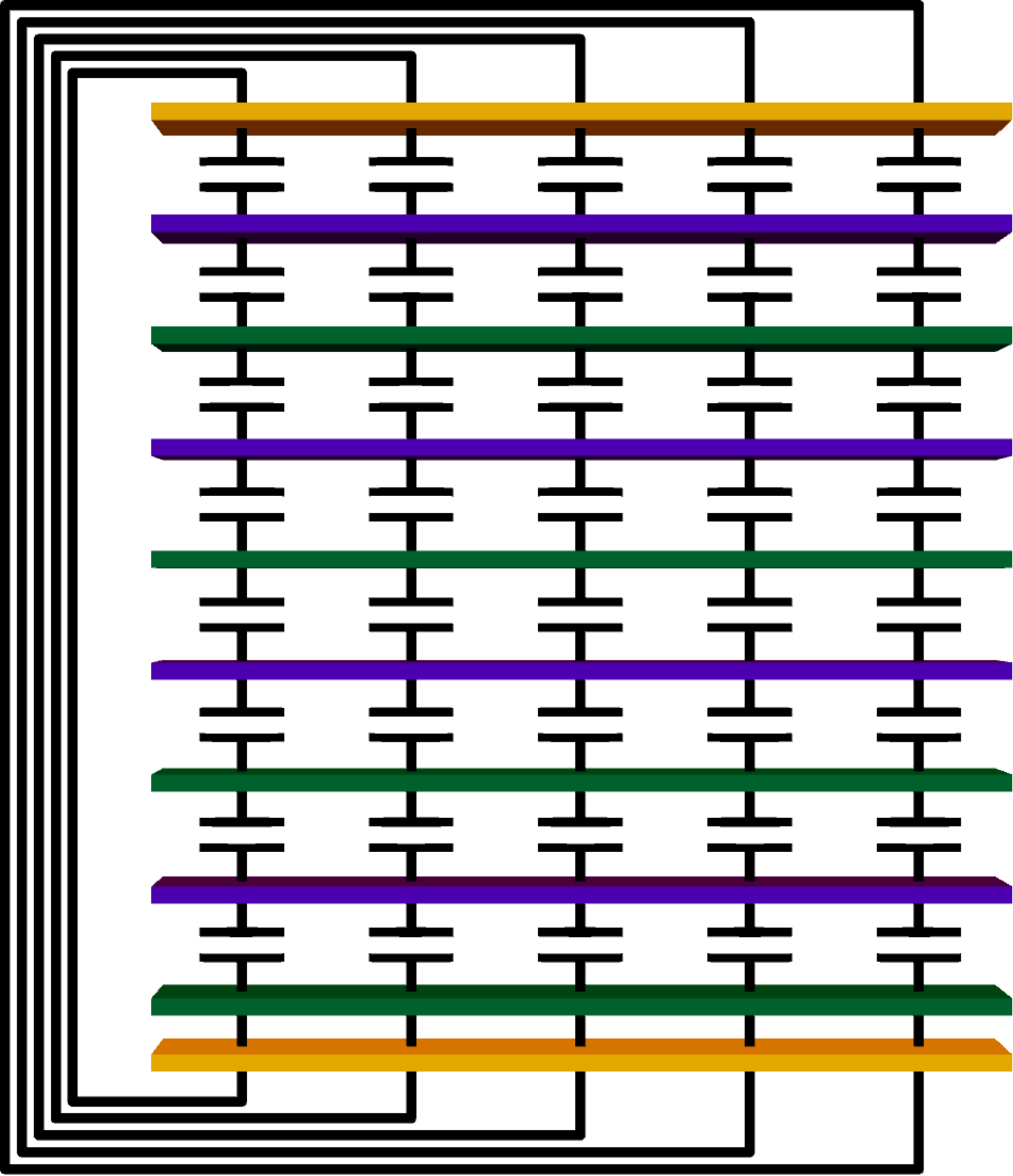}
\caption{\label{fig:boundary_z} Schematic of method of periodic boundary condition in the z dimension. Corresponding inductors in each LC circuit board (green or blue) are capacitively coupled to neighboring boards (These capacitors are actually installed on the boards, but we draw them in between boards for illustrative purposes - for similar reasons we depict 5 capacitors instead of the actual 16 apparent in the \emph{actual} boards). The two boards on the top and bottom (yellow) gather all the connections on all the inductors of the top and bottom circuit boards and wraps them around the side via four $2\times 32$ cable bands, realizing periodic boundary condition (PBC) in the z direction.}
\end{figure}

\subsection{Realizing Different Boundary Conditions}
\label{SI:RealizeBoundary}

The periodic boundary condition in the $\hat{u}$ dimension is printed onto the PCBs themselves. In the $\hat{v}$ direction, we design the board to leave three outlet connectors that enable us to choose between one of three boundary conditions (As illustrated in Fig.~\ref{fig:boundary_v}):
\begin{enumerate}
\item periodic boundary condition;
\item open boundary condition with 16 layers (full width);
\item open boundary condition with 15 layers
\end{enumerate}

A subtle aspect of the open boundary condition scenario is that one must be aware of the ``shifting term'' added onto the diagonal terms in the ``Hamiltonian'' caused by nearest-neighbor couplings. When we cut off the connection between the bottom and top layers of inductors we must introduce an additional coupling to compensate for the reduction in the ``shifting term''. This is achieved by adding two additional parallel capacitors onto the bottom layer of inductors, which are linked into the circuit in the second type of boundary condition (open boundary with 16 layers). In the third type of boundary condition (open boundary with 15 layers), the $16^{\text{th}}$ layer is shorted out, and thus the original connection-capacitors between the $15^{\text{th}}$ and $16^{\text{th}}$ layer become the compensating capacitors for the $15^{\text{th}}$ layer.

The periodic boundary condition in the $z$ dimension, which is perpendicular to the boards, is fulfilled by adding two connector boards above/below the top/bottom of the stack of boards, respectively, with 4 long cable bands connecting the two boards inducing the top-to-bottom coupling in the $z$-direction.

The validity of all of these boundary connections are guaranteed by the long wavelength of the RF field, which is several orders of magnitude larger than the scale of our boards, rendering negligible the phase difference between two ends of a cable.

\subsection{Probing System}
\label{SI:Probing_system}
As shown in Fig.~\ref{fig:whole_board}, the probing system consists of four parts: the board setup (the circuit boards, the 3D translational stage and the probe), the network analyzer, the time-resolved measurement unit (a function generator and an oscilloscope), and digital control unit (a Raspberry Pi, an Arduino, a two channel TTL switch and a Radiall 10-way switch).

In the main text, all the experimental data is collected in the frequency domain and thus does not require data from the time-resolved measurement unit. We probe the lattice spectrum using an RF network analyzer. The network analyzer's output channel, sweeping in the $200$kHz - $500$kHz frequency range, excites the lattice by magnetically driving a single lattice site(inductor) via one of the 10 pre-installed in-put coils as shown in Fig~\ref{fig:whole_board}(c); the network analyzer's input channel measures the response signal from a magnetic pick-up coil that is translated site-to-site in the lattice, on the end of a thin probe using a heavily modified 3D printer as the 3D translational stage, as shown in the lower-right corner of Fig~\ref{fig:whole_board}(d).

The entire measurement process consists of the following steps: 
\begin{figure}[H]
\centering
\includegraphics[width=0.42\textwidth]{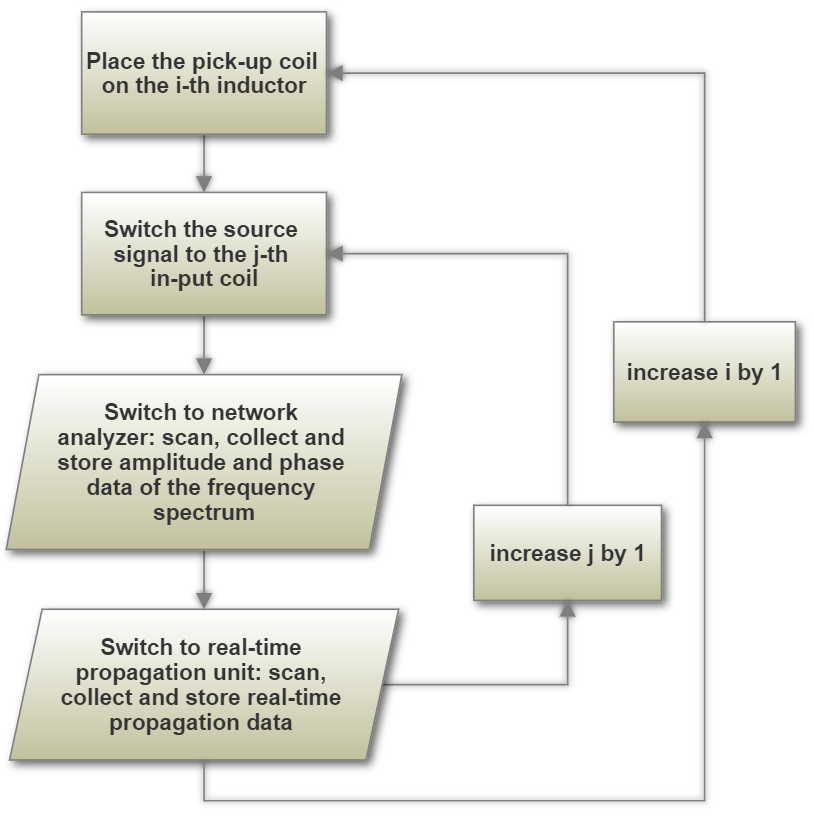}
\end{figure}
\begin{figure}
\centering
\includegraphics[width=0.49\textwidth]{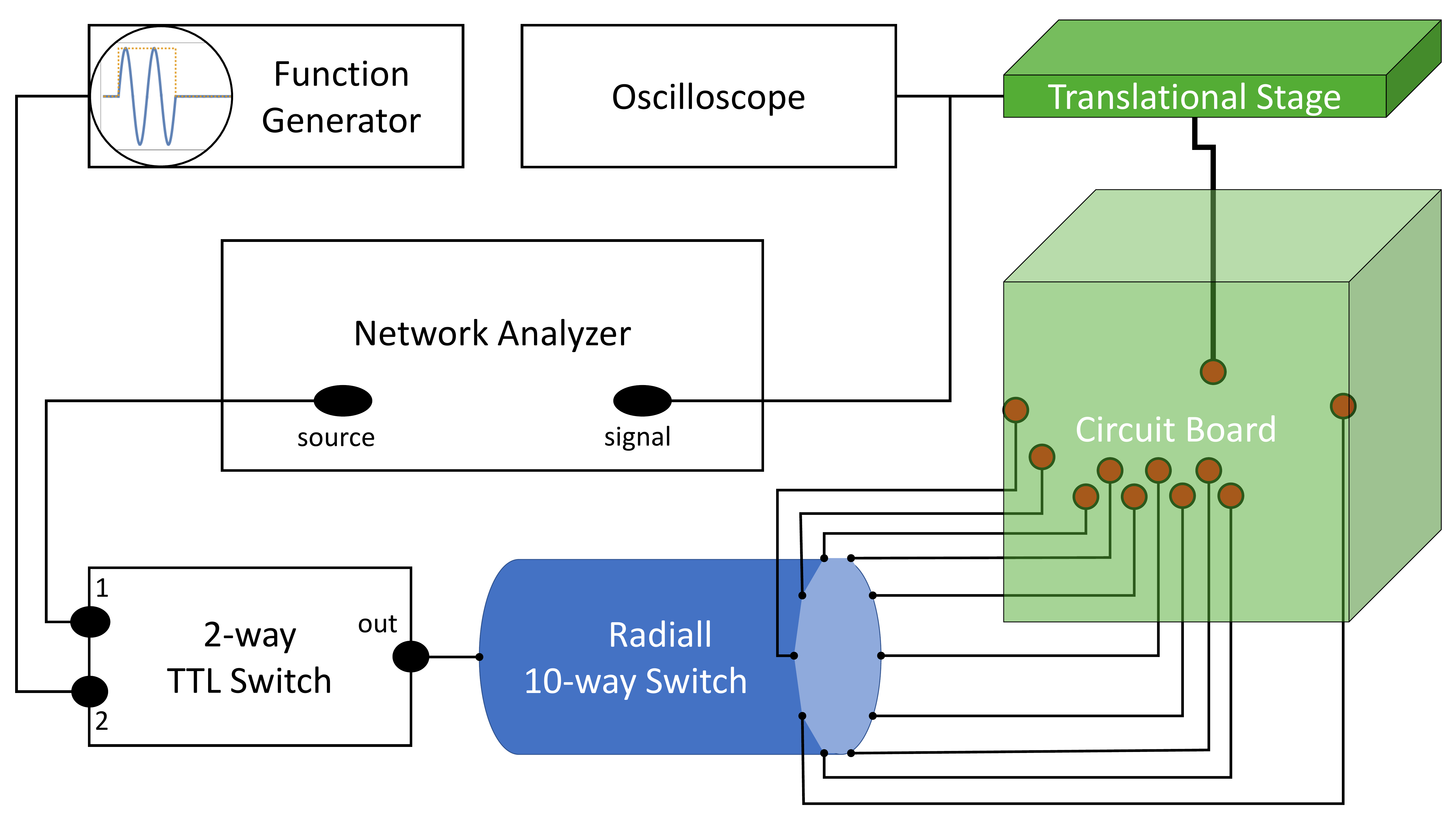}
\caption{\label{fig:probe_diagram}
A simple diagram of the entire measurement set-up. The small red circles inside the circuit board are the pick-up coil(wiring upward) and input coils (wiring downward).
}
\end{figure}

In Fig~\ref{fig:probe_diagram}, the Radiall 10-way switch (controlled by an Arduino) connects the source signal to one of the 10 input coils (we pick out 10 out of the 1024 sites where we install input coils around the inductors); the translation stage moves the pick-up coil to one of the 1024 sites; the 2-way TTL switch switches the measurement unit from the network analyzer to the time-resolved measurement unit. All above steps are controlled and coordinated by a server running on the Raspberry Pi.

\section{SPECTRAL FUNCTION, BULK BAND STRUCTURE AND SPIN TEXTURE}
\label{SI:BandSpinBerry}
To probe the bulk band-structure uncontaminated by any surface physics, we set up periodic boundary conditions on all surfaces. We then extract the band-structure and spin-texture of the Weyl circuit by measuring the spectral function in real-space, achieved by inducing an electric current on one of the inductors (e.g. $u=u_0, v=v_0, z=z_0, \sigma'=\uparrow \mathrm{or} \downarrow$) using an input coil (the source). We then measure the response function $P^{\sigma'}_{u,v,z,\sigma}(\omega)$ at all the 1024 sites over the frequency range of interest with a pick-up coil (the probe). It should be noted that $P^{\sigma'}_{uvz\sigma}(\omega)$ is complex, with magnitude and phase information thanks to time-resolution, compared to typical measurements in condensed matter or high-energy physics that only reflect the magnitude. Taking a discrete Fourier transform over all three spacial dimensions of the two-point function, we can view the excitation and the measurement process as: 1. The localized drive/input excites all points in momentum space with equal amplitude and definite pseudo-spin;  2. The  Fourier-transformed response function, selects out the states (in momentum space) that fall on the bands as a function of frequency, with (on each $\bm{k}$) a magnitude proportional to the projection of input signal's Bloch vector onto the Bloch vector of the eigenstate of the band.

\begin{figure*}
\centering
\includegraphics[width=0.8\textwidth]{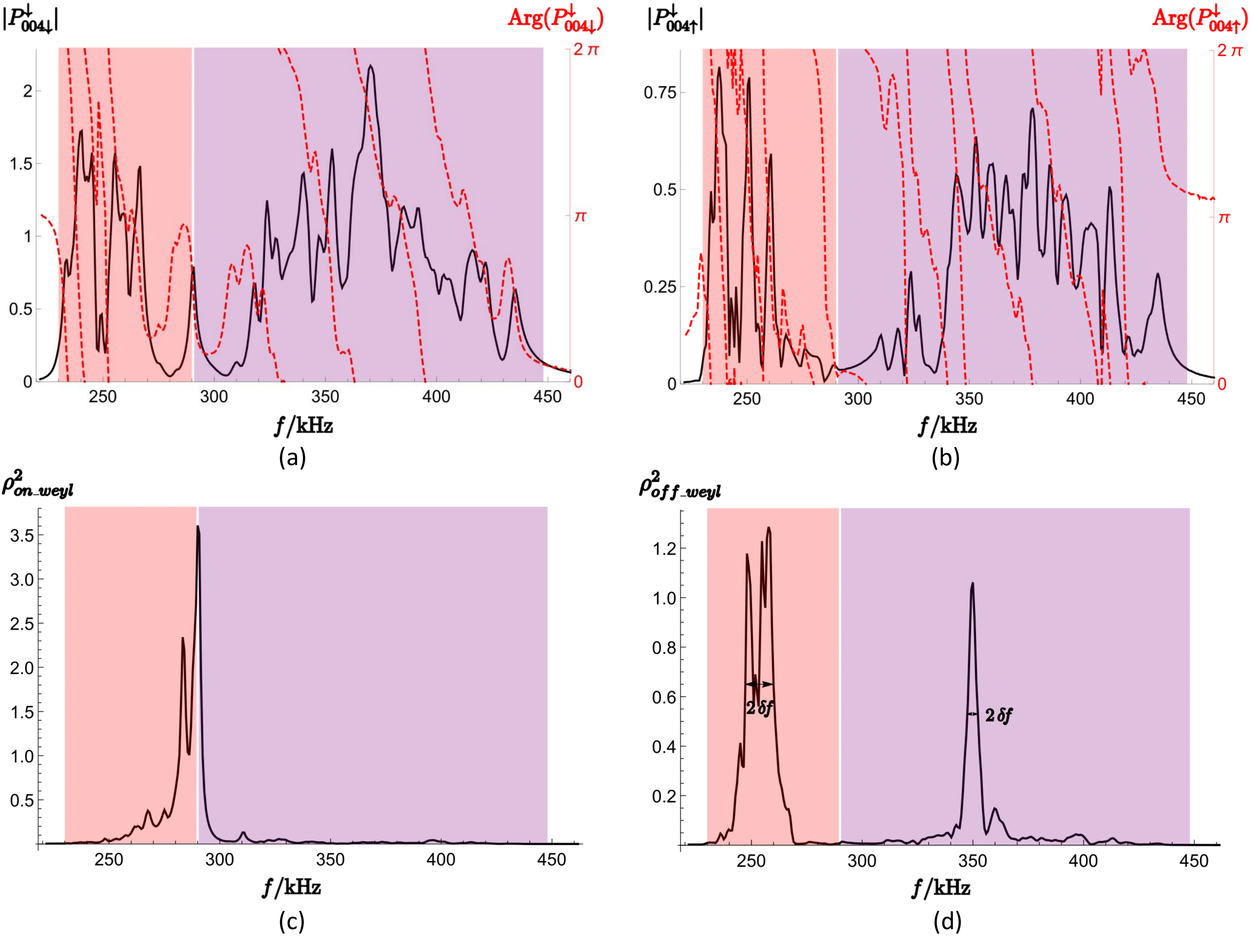}
\caption{\label{fig:NWA} Sample two-pointed spectra measured directly in real-space, and then converted to k-space. The two bands are in the frequency range marked with red $[231\,\text{kHz}, 290\,\text{kHz}]$, and purple $[290\,\text{kHz}, 447\,\text{kHz}]$, accordingly. \textbf{(a)}/\textbf{(b)} shows the \emph{spin down-spin down}/\emph{spin down-spin up} response function between two sites of the \emph{same/opposite} spin 4 sites apart in the $z$ direction, with magnitude depicted as a black solid line and phase as a red dashed line. In real space, we observe a separation in response into upper and lower bands, with a suppressed response near the Weyl energy and a small peak precisely \emph{at} the Weyl energy in the same-spin response \textbf{(a)}; this is because the Weyl nodes are spin-degenerate and therefore, exciting a spin-up mode will result in no response in the spin-down channel. \textbf{(c)}/\textbf{(d)} shows the spectral function in k-space \emph{at/away from} a Weyl node at the point $(-\pi/2,0,\pi/2)$/$(-\pi/2,0,0)$. A single peak at the Weyl frequency is observed at the Weyl point at $\textbf{k}=(-\pi/2,0,\pi/2)$ in \textbf{(c)} - which actually shows a weak splitting due to experimental disorder. By comparison, two peaks, one in each of the lower and upper bands, are observed away from the Weyl point at $\textbf{k}=(-\pi/2,0,0)$ in \textbf{(d)}.}
\end{figure*}

Experimentally, we impose periodic boundary conditions in all three dimensions; we then choose  10 sites distributed throughout both the surface and the bulk of the boards as source sites. We install an input coil around the inductor on each site, and connect the coils to the output of the network analyzer via a 10-channel Radiall 10 way switch. (At this point, two source sites with one on each sub-lattice are sufficient for bulk measurements by virtue of translational invariance. Only later on would we need more input channels to measure the surface modes in the slab geometry, as translational invariance is broken in the $\hat{u}$ direction.) The measurement probe, a small coil which is connected to the input channel of the network analyzer, is fixed onto a 3D translational stage that positions the probe on top of \emph{each and every} inductor to pick up the spin-and-amplitude-resolved response signal, $ P^{\sigma'}_{u,v,z,\sigma}(\omega) = A^{\sigma'}_{u,v,z,\sigma}(\omega) \exp\left(i \phi^{\sigma'}_{u,v,z,\sigma}(\omega)\right)$; here $\sigma'$ is the pseudo-spin of the input site, and $u,v,z$ and $\sigma$ are the relative spatial coordinates and pseudo-spin of the measured site, as illustrated in Fig.~\ref{fig:NWA} \textbf{(a)}/\textbf{(b)}. From these experimental data we calculate the spectral function and the spin texture in k-space (more precisely, on $8\times 8\times 8=512$ discrete points in k-space).

To extract the band structure information in k-space, we apply a discrete Fourier transformation to the response function:
$\tilde{P}^{\sigma'}_{\bm{k},\sigma}(\omega) =\tilde{P}^{\sigma'}_{k_u,k_v,k_z,\sigma}(\omega) = \frac{1}{8^3}\sum_{u,v,z=1}^{8} P^{\sigma'}_{uvz\sigma}(\omega) e^{i\left[\sqrt{2}(u-1)k_u+\sqrt{2}(v-1)k_v+(z-1)k_z\right]}$

\subsection{The Spectral Function}
\label{SI:Spectral_function}
Summing over the pseudo-spin degree of freedom, we get: 
$$\rho_{\bm{k}}(\omega)=\sum_{\sigma=\uparrow,\downarrow}\sum_{\sigma'=\uparrow,\downarrow}\left|\tilde{P}^{\sigma'}_{\bm{k},\sigma}(\omega)\right|^2$$

In Fig~\ref{fig:NWA} \textbf{(c)}/\textbf{(d)}, a comparison is drawn between the spectral function at fixed $\bm{k}$ at and away from a Weyl point. In Fig~\ref{fig:bandstructure}a, we plot spectral functions on points around a single Weyl point in 4D (3D wave-vector and a 1D energy), aligning multiple energy slices of the 3D density plot with separation that is proportional to the energy-shift. It is apparently that the dispersion relation around the Weyl point follows a linear conical form. 

\subsection{Bulk Band Structure}
\label{SI:Bulk_band}
At each point in k-space, the spectral function exhibits either one or two resonances that reflect the intersection-frequencies with the upper and lower bands, as shown in Fig.~\ref{fig:theory_vs_experiment}.

Experimentally, we do observe the dual-peaked spectrum at all points in k-space \emph{excluding} the 4 Weyl points, where the spectral function exhibits a single peak at $\omega=\omega_0$. We demarcate the frequency of these two bands in the k-space reciprocal lattice as: 
$\omega^{hi}_{k_u,k_v,k_z}$ and $\omega^{lo}_{k_u,k_v,k_z}$ for the upper band and lower band, respectively. Using polynomial interpolation, we extend the domain to the full Brillouin zone: 
\begin{align*}
\omega^{hi}_{\bm{k}}\xrightarrow{\text{interpolation}}\omega^{hi}(\bm{k}) \\
\omega^{lo}_{\bm{k}}\xrightarrow{\text{interpolation}}\omega^{lo}(\bm{k})
\end{align*}
Equi-energy surfaces are surfaces in k-space that satisfy $\mathcal{\omega}^{hi}(\bm{k})=\rm{const}$ or $\mathcal{\omega}^{lo}(\bm{k})=\rm{const}$; they are theoretically anticipated to expand spherically around the Weyl points near the Weyl frequency. 
\begin{figure*}
\centering
\includegraphics[width=\textwidth]{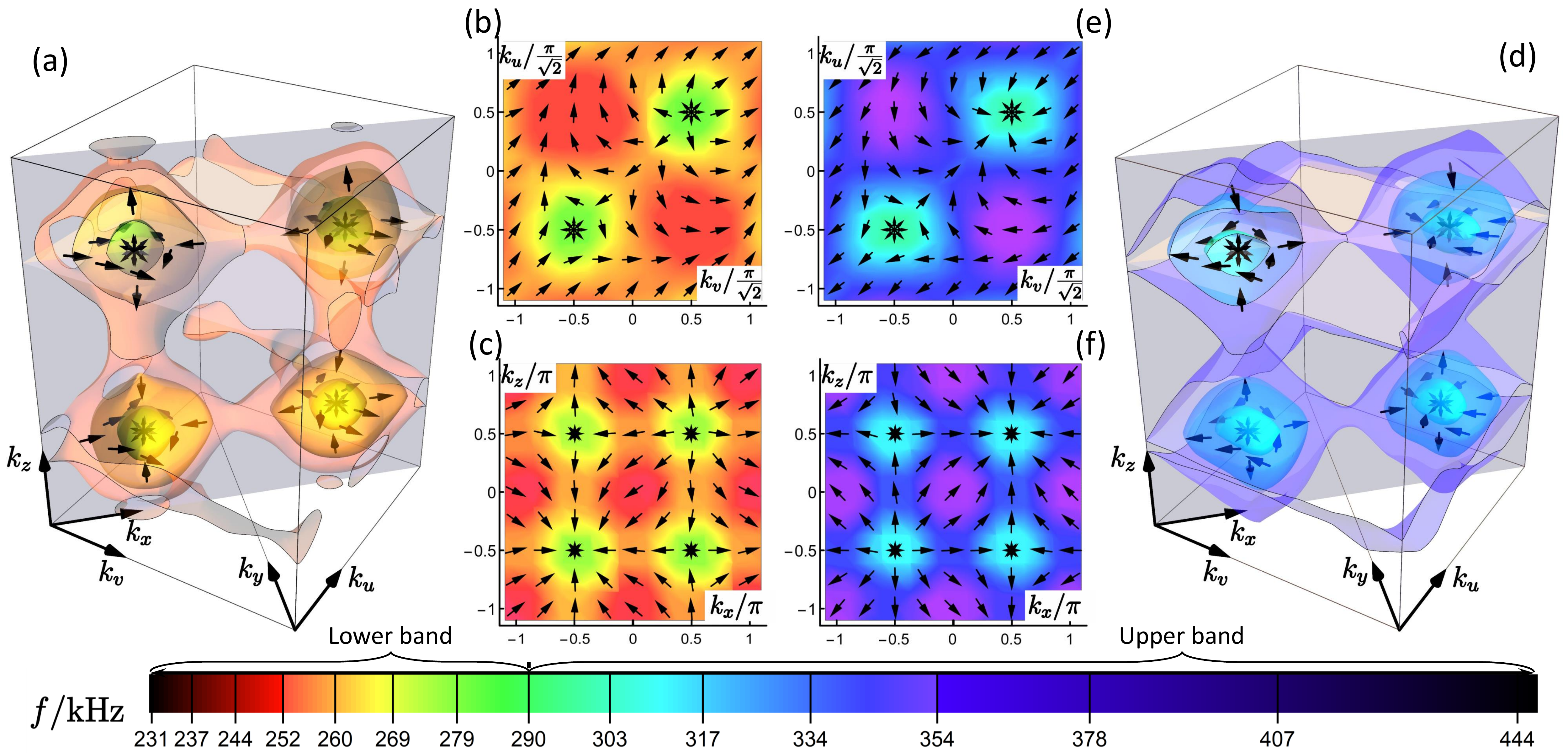}
\caption{\label{fig:SI_spin}  Spin-texture and equi-energy surface plots of both the lower- and upper- bands, as an extension of Fig.~\ref{fig:bandstructure} b, c and d. \textbf{(a)}, \textbf{(b)} and \textbf{(c)} are the color-coded equi-energy surfaces/heat map and the spin texture over the full 3D Brillouin zone, $k_z=\pi/2$ plane and $k_y=0$ plane of the lower frequency band, respectively. \textbf{(d)}, \textbf{(e)} and \textbf{(f)} are the colored equi-energy surfaces/heat map and the spin texture in the whole 3D Brillouin zone, $k_z=\pi/2$ plane and $k_y=0$ plane of the upper frequency band, accordingly. From the frequency-color-bar on the bottom one may infer the conical dispersion near the four Weyl nodes, that goes below/above the Weyl frequency of $\approx 290$ kHz in the lower/upper band. Also apparent are hedgehog/parabolic hedgehog spin textures around the Weyl nodes in the two bands.}
\end{figure*}

\subsection{Spin Texture}
\label{SI:Spin_texture}

We can quantify the spin texture of a state in k-space as the spinor:
$$\chi_{\bm{k}}(\omega)=\frac{\left(
   \begin{tabular}{c}
   $\tilde{P}^{\uparrow}_{\bm{k},\uparrow}(\omega)$  \\
   $\tilde{P}^{\uparrow}_{\bm{k},\downarrow}(\omega)$ 
   \end{tabular}
 \right)}{\norm{
   \begin{tabular}{c}
   $\tilde{P}^{\uparrow}_{\bm{k},\uparrow}(\omega)$  \\
   $\tilde{P}^{\uparrow}_{\bm{k},\downarrow}(\omega)$ 
   \end{tabular}
 }}
 =\frac{\left(
   \begin{tabular}{c}
   $\tilde{P}^{\downarrow}_{\bm{k},\uparrow}(\omega)$  \\
   $\tilde{P}^{\downarrow}_{\bm{k},\downarrow}(\omega)$ 
   \end{tabular}
 \right)}{\norm{
   \begin{tabular}{c}
   $\tilde{P}^{\downarrow}_{\bm{k},\uparrow}(\omega)$  \\
   $\tilde{P}^{\downarrow}_{\bm{k},\downarrow}(\omega)$ 
   \end{tabular}
 }}$$ 

The the second and third pieces of the equation should theoretically yield the same spin unless there is a degeneracy. An important exception occurs at the Weyl point, where the second and third pieces of the equation yield completely opposite spins - one points to the north pole on the Bloch sphere and the other south pole, both with large amplitude that is far larger than experimental error. This tells us that spin states are doubly-degenerate at the Weyl points, making that every spin state an energy-eigenstate (In Fig.~\ref{fig:bandstructure} \textbf{(b)} the doubly-degenerate spins are shown as a collection of arrows pointing in all directions). In other words, the response signal is anticipated to have the same spin state as the drive at these points, whereas for a non-degenerate point the spin and momentum are locked, with the amplitude of the response reflecting the projection of the drive onto the eigen-spin. 

More intuitively, the spin is denoted by the Bloch vector $\bm{a}$ - the expectation value of the vector spin operator $\bm{\sigma}$ operating on the state: \begin{align*}
a^x_{\bm{k}}(\omega) &=\langle\sigma^x\rangle_{\bm{k}}(\omega) =\sin(\theta)\cos(\phi) \\
a^y_{\bm{k}}(\omega) &=\langle\sigma^y\rangle_{\bm{k}}(\omega)=\sin(\theta)\sin(\phi) \\
a^z_{\bm{k}}(\omega) &=\langle\sigma^z\rangle_{\bm{k}}(\omega) =\cos(\theta)
\end{align*}
Where:
\begin{align*}
\theta &=2\arccot(\lvert\mathcal{R}_{\bm{k}}(\omega)\rvert)\\
\phi &=-\Arg \left(\mathcal{R}_{\bm{k}}(\omega)\right)
\end{align*}
Here $\mathcal{R}$ is the ratio between the first and second component of the spin spinor, $\mathcal{R}=\chi_\uparrow/\chi_\downarrow$.

Subscripts and arguments are suppressed for clarity. 

We are only concerned about the spin texture of the on-resonant states (the response at energies that fall on the dispersion curve at the chosen momentum), i.e. 
\begin{align*}
\bm{a}^{hi}_{\bm{k}}&=\bm{a}_{\bm{k}}(\omega^{hi}_{\bm{k}})\\
\bm{a}^{lo}_{\bm{k}}&=\bm{a}_{\bm{k}}(\omega^{lo}_{\bm{k}})
\end{align*}

The selected spin-texture of the upper band (that goes in the same direction as $\bm{h}(\bm{k})$) in the neighborhood of the four Weyl points is plotted in Fig.~\ref{fig:bandstructure} \textbf{(b)}, along with the equi-energy surfaces. We see that points $(k_x,k_y,k_z)=(\pi/2,0,\pi/2)$ or $(-\pi/2,0,-\pi/2)$ have negative chirality, and points $(\pi/2,0,-\pi/2)$ and $(-\pi/2,0,\pi/2)$  have positive chirality. We also plot the full spin-texture of the upper band on two selected planes in k-space in Fig.~\ref{fig:bandstructure} \textbf{(c)}-\textbf{(d)}. We show the spin-texture and the equi-energy surfaces of both the upper and lower band in Fig.~\ref{fig:SI_spin}.

\section{BERRY CURVATURE AND TOPOLOGICAL INVARIANTS}
\label{SI:BCandTI}

\begin{figure*}
\centering
\includegraphics[width=\textwidth]{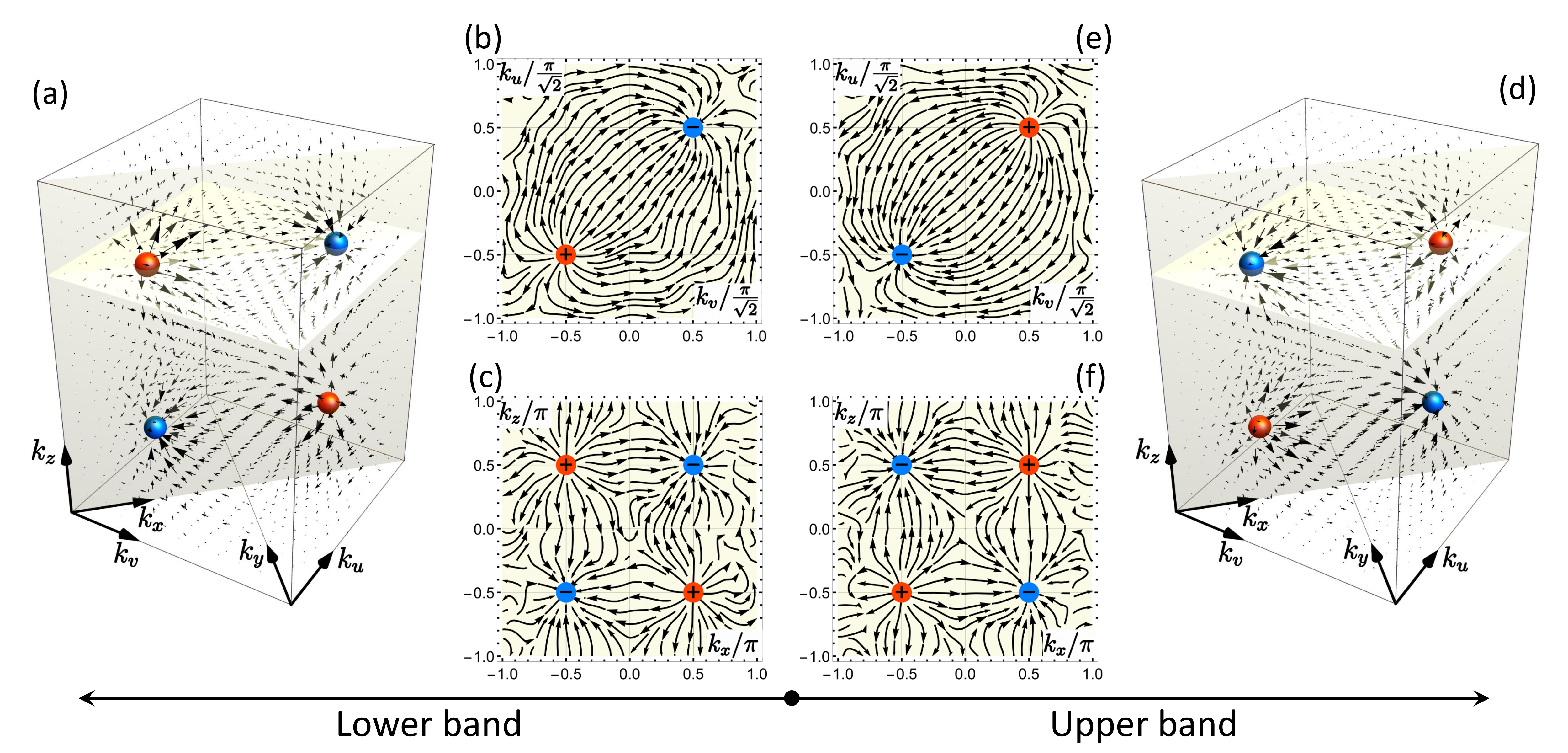}
\caption{\label{fig:SI_Berry} \textbf{Berry Curvature and Chiral charge of both bands}. In \textbf{(a)}, \textbf{(b)} and \textbf{(c)}, we observe the Berry curvature flow and the chirality of the Weyl nodes (orange for positive charge and blue for negative charge), which is the same as in Fig.~\ref{fig:curvandsurf}a,b and c. In In \textbf{(d)}, \textbf{(e)} and \textbf{(cf)}, we see the calculated Berry's curvature of the upper band which swaps sources and sinks compared to the lower band. By convention we define the Berry's curvature and the chiral charge of the lower band to be the Berry curvature and the chiral charge of the material. One may also employ the curvature of the upper band, which is in theory strictly opposite to the curvature of the lower band, and offers higher quality data due to its wider bandwidth.}
\end{figure*}

\subsection{Berry Curvature}
\label{SI:BerryCurvature}

In a two-band tight-binding model, the Berry's curvature pseudo-vector is defined as follow ($n=hi \,\mathrm{or}\, lo$ for band number): 
$$
\Omega^{n}(\bm{k})=\bm{\nabla_{k}}\times \bm{A}^{n}(\bm{k})
$$
$\bm{A}^{n}(\bm{k})$ is the Berry's connection, 
$$
\bm{A}^{n}(\bm{k})=
\mel{\psi^n(\bm{k})}{i \bm{\nabla_{k}}}{\psi^n(\bm{k})}
$$
Which gives us,
\begin{equation}
\label{SI_Eqn:BerryCurvature}
\Omega^{n}(\bm{k})=i 
\mel{\bm{\nabla_{k}} \psi^n(\bm{k})}{\times}{\bm{\nabla_{k}}\psi^n(\bm{k})}
\end{equation}

We further write the wavefunction as a product of spatial- and spinor- comonents. One can prove that in tight-binding model, the spatial wave-function does not contribute to the Berry curvature~\cite{He2012}. Therefore, we rewrite the expression as:
\begin{align*}
\Omega^{n}(\bm{k})&=i \left(
\bm{\nabla_{k}} \widetilde{\chi^{n*}}(\bm{k})\right)
\times
\left(\bm{\nabla_{k}}\chi^{n}(\bm{k})\right)
\\
&=2\sum_{\alpha=\uparrow,\downarrow} \left(
\bm{\nabla_{k}}\Im{\chi^n_\alpha(\bm{k})}\right)
\times
\left(\bm{\nabla_{k}}\Re{\chi^n_\alpha(\bm{k})}\right)
\end{align*}
Where $\chi^n(\bm{k})$ is defined on the continuous Brillouin zone as the ``interpolation function'' (\emph{This is not simple interpolation}, for details see SI~\ref{SI:Spin_interpolation}) of $\chi^n_{\bm{k}}$ values on discrete points in k-space.  
There is not a single gauge one can choose to allow $\chi(\bm{k})$ to be continuous (although $\bm{a}(\bm{k})$ is indeed continuous) on the entire Bloch sphere, but we can get around this by choosing the convention $\chi=\left(
   \begin{tabular}{c}
  $\cos\left(\theta/2\right)$  \\
  $e^{i\phi}\sin\left(\theta/2\right)$ 
   \end{tabular}
 \right)$ on the northern hemisphere and $\chi=\left(
   \begin{tabular}{c}
  $e^{-i\phi}\cos\left(\theta/2\right)$  \\
  $\sin\left(\theta/2\right)$ 
   \end{tabular}
 \right)$ on the southern hemisphere. 

By convention, the Berry's curvature of a two-band material is defined on the lower band, as it is the occupied band for fermions. The results are shown in Fig.~\ref{fig:curvandsurf}a. As an illustration of the lower band, We can see Berry's flux emerging from the two positive chiral charges, whose divergence of curvature is akin to that of an point electric charge as shown in orange, to the two negative chiral points as shown in blue. Fig.~\ref{fig:curvandsurf}b-c are two 2D cross sections of the 3D Brillouin zone whose streamlines depict the Berry curvature's flow. 

\subsection{Spin-Interpolation Technique} 
\label{SI:Spin_interpolation}

We interpolate the 3D Bloch vector over points on the Brillion zone except the doubly degenerate points that sit at the Weyl nodes. The reason we interpolate the Bloch vector instead of the 2 component spinor is that: (1) the mapping from the Bloch sphere to the spinor is not continuous over the entire sphere due to the chiral charge; there is at least one singularity point. (2) We cannot express the doubly degenerate state of the Weyl point in spinor form, but in the Bloch representation it is essentially a mixed state located in the center of the Bloch sphere. 

The spirit of spin interpolation is therefore that we interpolate in Bloch-vector space and then transform the interpolation function back into a spinor with a gauge that is continuous in our region of interest; we employ this interpolated spinor to calculate the Berry's curvature according to Eqn.~\ref{SI_Eqn:BerryCurvature}. 

We take the lower band for example (upper band is the same): First we have the Bloch vector for 8*8*8=512 sample points states in the k-space $\bm{a}^{lo}_{\bm{k}}$. They are measured definitively at all points except the four points where degeneracy occurs. These 4 points are: $(k_u,k_v,k_z)=\left(\frac{\pi}{2\sqrt{2}},\frac{\pi}{2\sqrt{2}},\frac{\pi}{2}\right), \left(\frac{\pi}{2\sqrt{2}},\frac{\pi}{2\sqrt{2}},-\frac{\pi}{2}\right), \left(-\frac{\pi}{2\sqrt{2}},-\frac{\pi}{2\sqrt{2}},\frac{\pi}{2}\right), \\ \left(-\frac{\pi}{2\sqrt{2}},-\frac{\pi}{2\sqrt{2}},-\frac{\pi}{2}\right) $, which we later confirm to be Weyl points. We denote them as $\bm{k}_i^{Weyl},\quad (i=1,2,3,4)$.
For these 4 points, because we know that they are doubly degenerate, we further let $\bm{a}^{lo}_{\bm{k}_i^{Weyl}}=0,\quad (i=1,2,3,4)$. As a result, we have null Bloch vectors at the Weyl points but unit-length Bloch vectors at other points in k-space. Up to now, we have  a complete set of $\bm{a}^{lo}_{\bm{k}}$ defined on uniform lattice in k-space.

Next, we perform an interpolation over the 3D space embedding the Bloch vector to expand $\bm{a}^{lo}_{\bm{k}}$ to smoothly cover the full Brillouin zone. 
$$\bm{a}^{lo}_{\bm{k}}\xrightarrow{\text{interpolation}}\bm{a'}^{lo}(\bm{k})$$
Note that our n{\"a}ive interpolation process takes the Bloch vector off the surface and into the volume of Bloch sphere. Accordingly, we renormalize the interpolated Bloch vectors at all points in the BZ except at the four Weyl nodes. 

$$\bm{a'}^{lo}(\bm{k})\xrightarrow{\text{unification}}\bm{a}^{lo}(\bm{k})$$

Finally we reconstruct the spinor, and from there the Berry's curvature field, from the Bloch vector field. 

For any small region in the Brillouin zone that excludes the 4 Weyl points (On these singularity points the spinor and Berry's curvature are ill-defined), we make a gauge choice for the two component wavefunction as follows: (1) The Bloch vector function in a region excluding the south pole, i.e., $\theta\leq\pi/2$, we let $\chi^{lo}(\bm{k})=\left(
   \begin{tabular}{c}
  $\cos\left(\theta/2\right)$  \\
  $e^{i\phi}\sin\left(\theta/2\right)$ 
   \end{tabular}
 \right)$;  (2) The Bloch vector function in the region excluding the north pole, i.e., $\theta>\pi/2$, we let $\chi^{lo}(\bm{k})=\left(
   \begin{tabular}{c}
  $e^{-i\phi}\cos\left(\theta/2\right)$  \\
  $\sin\left(\theta/2\right)$ 
   \end{tabular}
 \right)$.
 Both gauges can be applied if the small region includes neither poles. This procedure enables us to avoid singularities on the surface of the Bloch sphere.

\subsection{Topological Invariants}
\label{SI:Topological_invariant}

Each Weyl node acts as either a source or sink of Berry curvature, as shown from flow of the curvature vectors into/out-of the nodes in~\ref{fig:SI_Berry}, where 2D slices of the full 3D Brillouin zone Berry curvature of both bands are displayed. The colored spheres in each plot show the sources and sinks of the Berry curvature $\bm{\Sigma^n(\bm{k})}$, which can also be understood as magnetic monopoles in momentum space located at the four Weyl nodes. The divergence of the Berry curvature field of the lower band gives us the chiral charge density $\rho_chi=\frac{1}{2\pi}\nabla_k\cdot\bm{\Omega^{lo}}$: two nodes exhibit positive charge (orange), and the other two exhibit negative charge (blue).

The Chern number of any closed surface enclosing a single Weyl node at $\bm{q}=\left(\pm \pi/2,0,\pm \pi/2\right)$, i.e. the Berry-flux through that surface, yields chiral charges of the enclosed Weyl node, $$\chi=\frac{1}{2\pi}\oiint \bm{\Omega} \cdot \hat{\bm{n}} \mathop{}\!\mathrm{d} S = - \,\text{sgn}(q_x)\, \text{sgn}(q_z)$$. When no singularity exists inside the surface, the Berry's connnection field is smooth and well-defined everywhere on the surface and thus the surface integral of Berry curvature vanish due to $\Omega^{n}(\bm{k})=\bm{\nabla_{k}}\times \bm{A}^{n}(\bm{k})$, analoguous to the vanishing surface integral of a sourceless magnetic field; on the other hand, if a singularity (Weyl node) lives in the volume enclosed by the surface, the Berry's connection can only be defined (in \emph{any} continuous gauge) everywhere except for a string that starts from one Weyl point and ends at its opposing chirality twin - this gives rise to a Berry curvature field with a point source, analogous to the $\bm{A}$ and $\bm{B}$ field around a magnetic monopole~\cite{dubvcek2015weyl,Bernevig2015}. However, according to the famous Gauss-Bonnet theorem, these surface integrals can only yield integer results, hence they are topological invariants known as the first Chern class topological invariant. This can also be understood through the analogy between closed 2D $\bm{q}$-space surfaces enclosing a Weyl node and bands of a 2D Chern insulator~\cite{hasan2017discovery}. The specific shape of the integrated surface is arbitrary so long as it is closed - this also includes those surfaces that exploit the toroidal topology of the Brillouin zone - such as the cylindrical surface in~\ref{SI:Surface_orientation}.

\section{SURFACE BAND STRUCTURE AND BULK-SURFACE CORRESPONDENCE}
\label{SI:SurfaceBulk}

In this section, we examine the bulk-surface correspondence band structure obtained in the slab geometry configuration of our circuit, which means periodic boundary conditions in the $u$ and $z$ directions, with open boundary conditions in the $v$ direction. In other words, the system is equivalent to a 16(15) layer slab that stretches to infinity in the two in-plane dimensions. The slab has thus exhibits two surfaces separated by a single bulk.

\subsection{Choice of surface orientation}
\label{SI:Surface_orientation}

\begin{figure}
\centering
\includegraphics[width=0.44\textwidth]{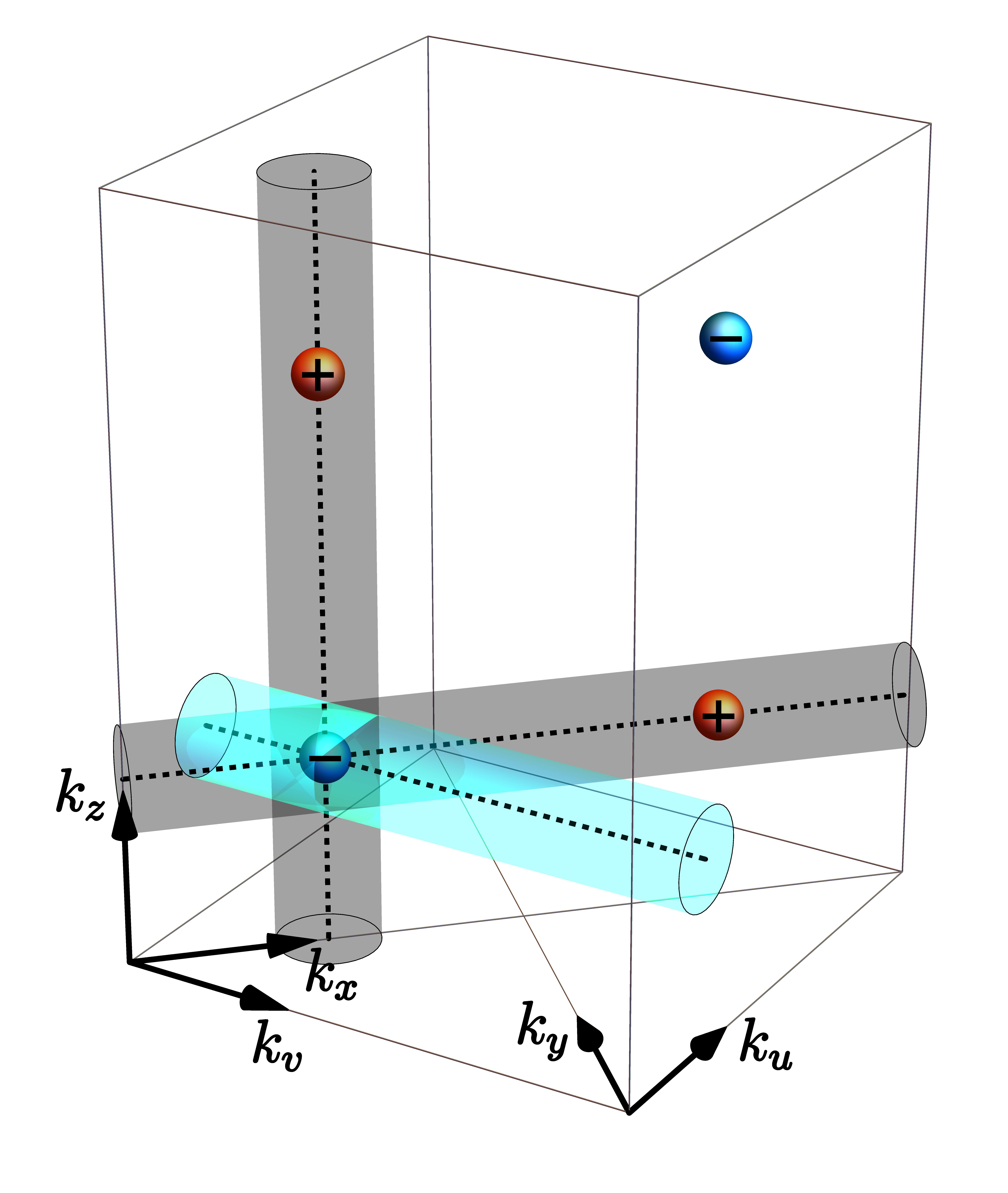}
\caption{\label{fig:Cylinders}
The sum of chiral charges of Weyl points projected onto a surface Brillouin zone determines whether the surface will exhibit Fermi arc surface states. Two Weyl nodes of the opposite chiral charge cancel one another when projected along the $k_z$/$k_x$ direction, as shown on the axes of the vertical/horizontal gray cylinder -- there are thus no Fermi arc surface states on $x-y$/$y-z$ surfaces. In our experiment we instead consider the $u-z$ surface: the $k_u-k_z$ BZ-surface exhibits a single projected Weyl node -- that is, the cylinders whose axes point in the $k_v$ direction enclose a single Weyl node and thus have a (non-zero) Chern number of $\pm 1$ (e.g. the blue cylinder in the figure). The system should thus exhibit topologically protected surface states on $u-z$ oriented surfaces.}
\end{figure}

We choose the $u-z$ oriented surface because not all surface orientations yield topologically protected Fermi arcs- only selected surfaces on which the projection of the chiral charges of the Weyl nodes do not cancel one another. We can understand the bulk-surface correspondence of $u-z$ oriented surfaces by computing the Chern number of a cylindrical surface in the BZ, whose axis intersects a positive/negative Weyl node in the $q_v$ direction (The blue cylinder shown in Fig.~\ref{fig:Cylinders}):

The Chern number of the cylindrical surface equals the sum of all enclosed chiral charges - if this sum is zero the 2D band-structure on the cylinder is trivial and exhibits no protected edge states, as in the case of the gray cylinders in Fig.~\ref{fig:Cylinders}; on the other hand, if the Chern number is \emph{non-zero}, then there are a pair of protected edge states that reside on the top and bottom edge of the cylinder, as in the  case of the blue cylinder in Fig.~\ref{fig:Cylinders}. As the radius of the blue cylinder expands from zero, an \emph{arc} will form in the surface band-structure, until it terminates when the cylinder encloses a Weyl point of opposite chirality, thus connecting the surface projection of the Weyl nodes of opposite chiralities. The existence of this ``Fermi Arc'' is often taken as the smoking-gun signature of the chirally-charged Weyl-nodes.

\begin{figure*}
\centering
\includegraphics[width=0.99\textwidth]{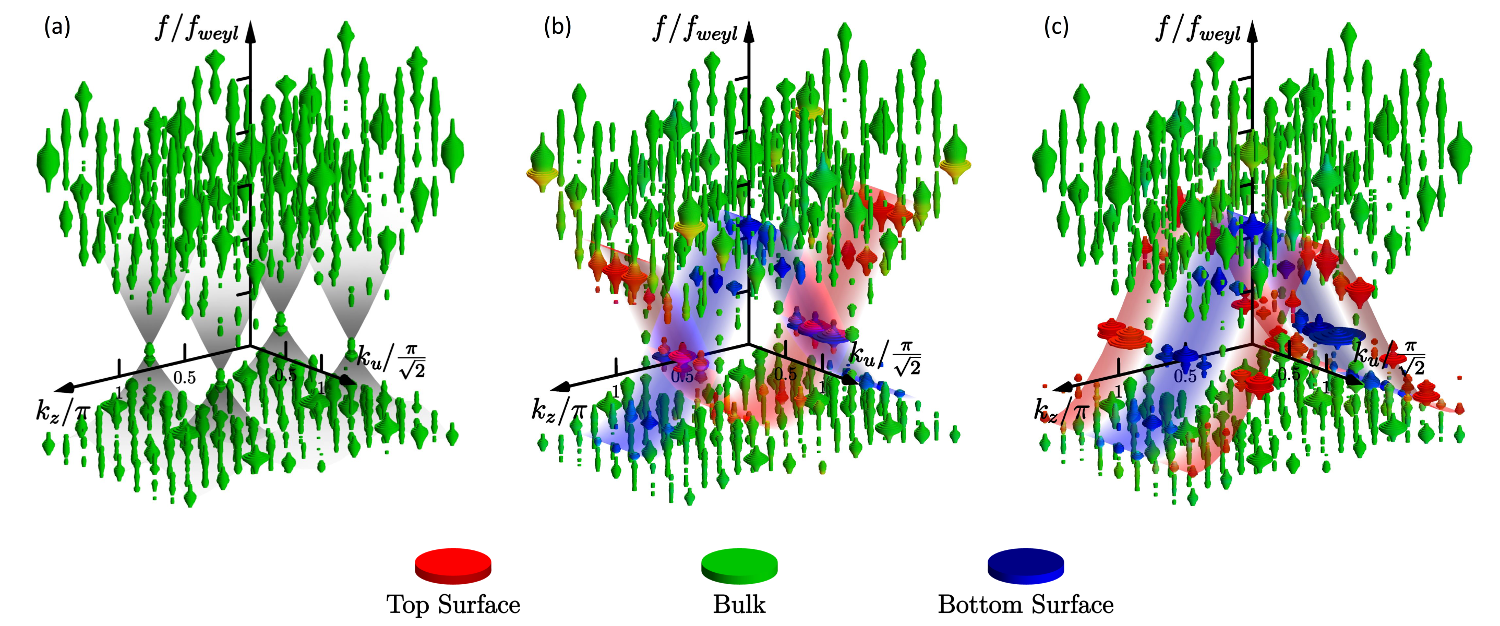}
    \caption{\label{fig:UZband} A comparison of the measured projected spectral functions in the $k_u-k_z$ plane. \textbf{(a)} is the all-bulk periodic-in-all-direction geometry, \textbf{(b)} is the 16-layer slab geometry and \textbf{(c)} is 15-layer slab geometry. The position of the disks marks the $k_u,\, k_z,\, E$ coordinates, their radii reflect the summed amplitude in the $k_v$ direction (akin to the density of states), and the color of the disks reflects how close to surface a mode is: the red channel indicates a top-surface mode; blue-channel a bottom-surface mode; and green-channel a bulk-penetrating mode. Accordingly, a mode which co-exists on both surfaces is purple. Blue, red and gray glow are guides to the eye, outlining of the bulk cones and surface bands.}
\end{figure*}

\subsection{Extracting the Surface States}
\label{Extracting_surface}

Similarly to the 3D periodic situation, we excite the lattice at 10 different sites and measure at all 1024 sites. We obtain the bulk spectrum by measuring bulk-bulk response. We then measure surface-surface response, which we convert to a surface band-structure via a 2D discrete Fourier transformation. 

Consider, for example, the top(bottom) surface: We measure the  surface spectral function by the following procedure: 

\begin{enumerate}
\item Select the channel of input on the top(bottom) layer, and measure at this and every other site of the same layer (64 sites in total). 
\item Apply a 2D discrete Fourier transformation on the measured response functions:

$\tilde{P}^{+(-)}_{k_u,k_z}(\omega) = \sum_{u,z=1}^{8} P^{+(-)}_{u,z}(\omega) e^{i\left[\sqrt{2}(u-1)k_u+(z-1)k_z\right]}$

\item The top(bottom) surface spectral function in the surface Brillouin zone is then:
$$\rho^{+(-)}_{k_u,k_z}(\omega)=\left|\tilde{P}^{+(-)}_{k_u,k_z}(\omega)\right|^2$$
\end{enumerate}

In Fig.~\ref{fig:curvandsurf}d\&e we plot the super-imposed surface bands in a two-color-scheme where red indicates top surface weight and blue bottom surface weight. In addition, we include bulk states as the sum over all inner layers: 
$$\rho^{bulk}_{k_u,k_z}(\omega)=\sum_{i=2}^{layer-1}\left|\tilde{P}^{i}_{k_u,k_z}(\omega)\right|^2$$
Where $\tilde{P}^{i}_{k_u,k_z}(\omega)$ is the Fourier transformed response function of the i-th layer. Super positioning $\rho^{+}_{k_u,k_z}(\omega)$, $\rho^{-}_{k_u,k_z}(\omega)$ and $\rho^{bulk}_{k_u,k_z}(\omega)$, we see the surface-vs-bulk band structure as shown in Fig.~\ref{fig:UZband}, where the radii of the disks represent the density of states, and the color of the disks marks the positioning of the mode by a three-color-scheme - red(blue) channel for top(bottom) surface, and green channel for the bulk.

\subsection{Slab geometry: 16 layers in the $\hat{v}$ direction}
\label{16_layers}

This is a representative situation for even-layered samples. We have 16 layers of crossed-stacked sites along the $\hat{u}$ direction, with the top layer (\# 1) composed of A-sites and the bottom layer (\# 2) composed of B-sites. 

In Fig.~\ref{fig:curvandsurf}d we plot several energy cuts of the of surface spectral function of both surfaces. The red channel reflects weight on the top surface, and the blue channel the bottom surface. The third slice in the center shows an overlap of top and bottom states in purple; this is the Fermi arc. 

Comparing these Fermi arcs to the projections of Weyl points onto the surface Brillouin zone, we see that the arcs connect a projection of positive chirality Weyl point to a projection of the negative chirality Weyl point on the top surface then trace back along the same line on the bottom surface, as predicted by theory. 

\subsection{Slab geometry: 15 layers in the $\hat{v}$ direction}
\label{SI:15_layers}
This is a typical situation for a sample with an odd-number of layers. Following the same procedure as the 16 layer case, we obtain the surface spectral functions in the 15-layer slab geometry shown in Fig.~\ref{fig:curvandsurf}e. The only difference is that the top surface (red) undergoes a gauge transformation and shifts half a Brillouin zone in both directions, giving a different form of Fermi arcs. These Fermi arcs connects the projection of positive and negative chirality Weyl points through the inside of the bottom surface Brillouin zone over $k_u=0$, and wrap around the outside of the top Brillouin zone over $k_u=\pm \sqrt{2}/2$.  The shape difference of Fermi arcs can be understood as a $(\pi,\pi)$ translation in momentum space that originates from a gauge transformation, $$U=\exp[i(\frac{\pi z}{a}+\frac{\pi u}{\sqrt{2} a}+\frac{\pi}{2}c_A^\dagger c_A)],$$ that maps between alternating layers in the $v$ direction, as shown in~\ref{fig:whole_board}a. However, in general, the shape of Fermi arcs on the top and bottom layer may form a closed loop when the number of layers is odd, and must trace back on themselves when the number of layers is even~\cite{Hosur2012}.

\section{TIME-RESOLVED DYNAMICS}
\label{SI:TimeDynamics}
In this section we measure the time-dynamics of a temporally short square pulse, spectrally centered on the Weyl frequency, injected into a single lattice site. We measure the time-delayed response at all other sites, and by analyzing the relationship between the position of the output site and the time of delay, we reconstruct the temporal propagation of the spherical wavefront.

\begin{figure}
\centering
\includegraphics[width=0.49\textwidth]{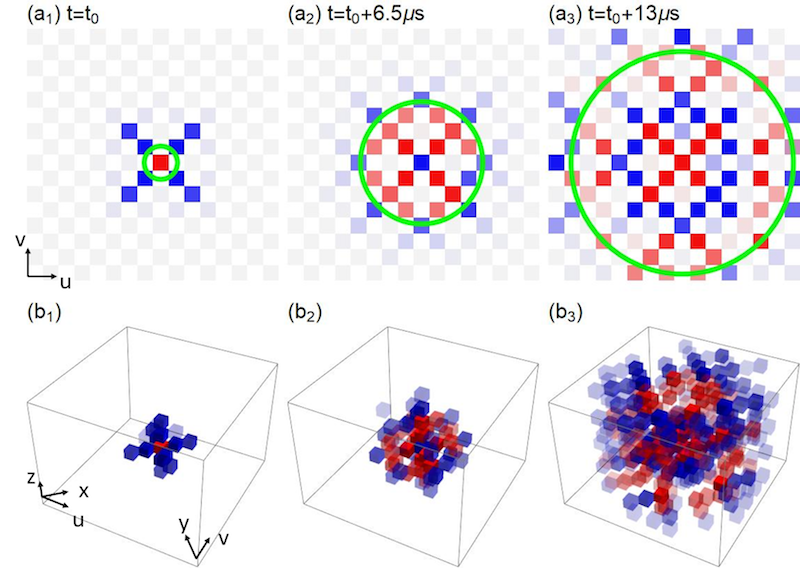}
\caption{\label{fig:expansion} \textbf{Wavefront Imaging Following Local Excitation.} After a square-pulse-modulated wavepacket at Weyl frequency is injected at single site, we measure time- and site- resolved signal intensity in the system. In the 2D cross-section containing the source site ($\mathbf{a_1}$ to $\mathbf{a_3}$) or the entire 3D system ($\mathbf{b_1}$ to $\mathbf{b_3}$), each site is represented as a light-gray square or a cuboid respectively. Cuboid opacity reflects signal intensity, while color indicates whether response amplitude is increasing (blue) or decreasing (red) at the measurement time. The first peak at each measurement time is emphasized with a green circular guide to the eye in the 2D cross-section. We thus observe the wave peak expanding through the system roughly spherically around the source site, with isotropic speed.}
\end{figure}

Driven near the Weyl frequency, the system is expected to exhibit a linear, isotropic dispersion. In our realization, a two-cycle square pulse at the Weyl frequency centered at $t=0\mu s$ (as shown in Fig~\ref{fig:probe_diagram}) is injected into a source site. Response is measured in-situ, and amplitude is extracted from the envelope of each raw oscilloscope time-trace. The intensities at three uniformly spaced times are presented in Fig.~\ref{fig:expansion}.

\begin{figure}
\centering
\includegraphics[width=0.47\textwidth]{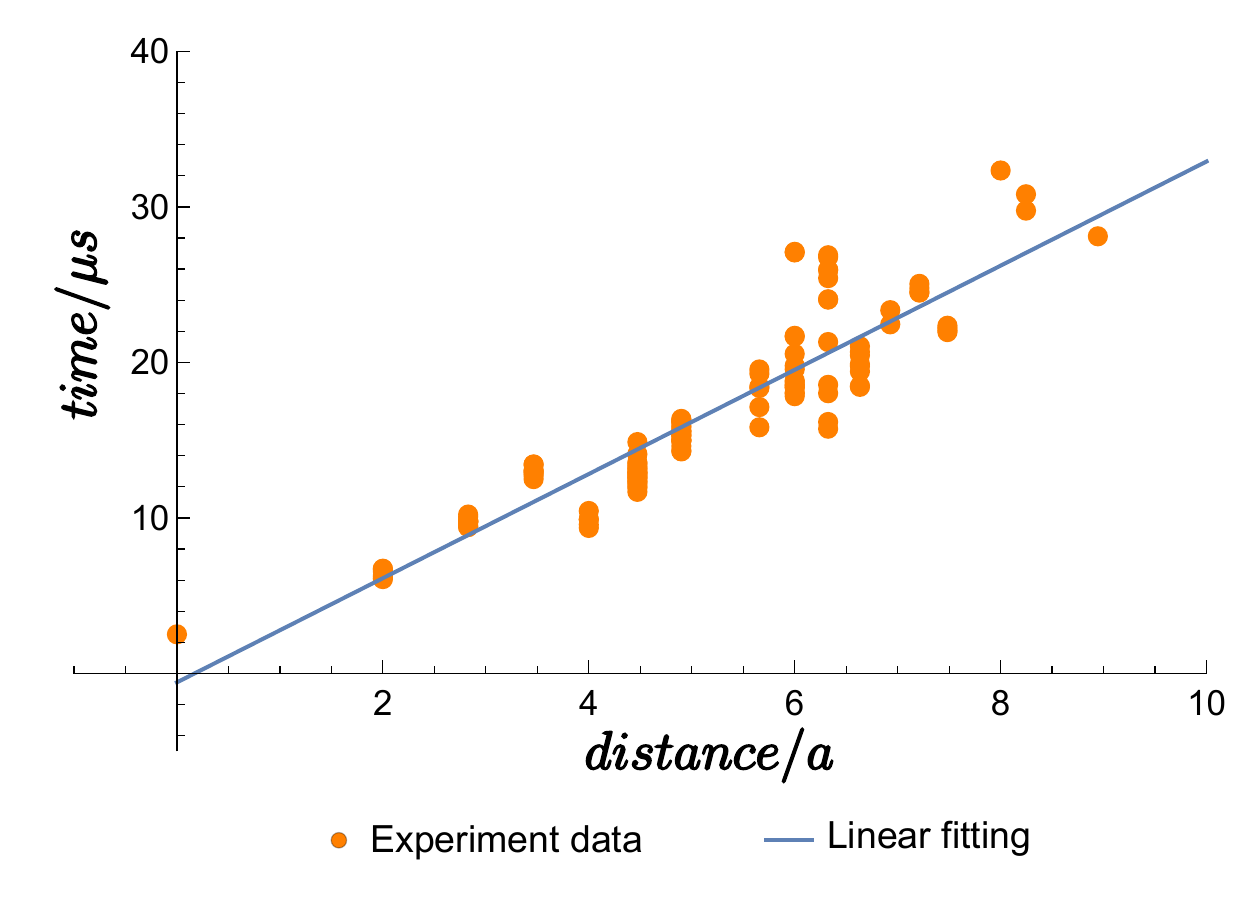}
\caption{\label{fig:propagationfit} \textbf{Pulse propagation near the Weyl point.} Peak of the response envelope as a function of distance from the source. The data are plotted in orange, and the best fit in blue. The inverse of the slope provides a measurement of the group velocity.}
\end{figure}

To extract the group velocity, we plot and fit in Fig~\ref{fig:propagationfit} the signal delay time versus the distance to point source on 128 of the 1024 sites, whose signal amplitudes are strong and less influenced by noise. By computing the inverse of the slope of the fitted line, we obtain a group velocity of 
$$v_{g}^{\text{experimental}}=299 \pm 11\,\, \text{ms}^{-1}$$

where we assign $a=1$ as the distance between nearest neighbor sites. This group velocity is consistent with the derived value from SI:~\ref{SI:WeylCircuitComponents}: 
\begin{align*}v_{g}^{\text{theoretical}}&= 2\pi \lvert \nabla_{\bm{k}}f_+ \rvert_{\bm{k}\rightarrow Weyl\, point}\\&=\frac{1}{6\sqrt{3 L_0 C_0}}=304 \,\,\text{ms}^{-1}\end{align*}

\end{document}